\documentclass[reprint,superscriptaddress,amsmath,amssymb,floatfix,aps,footnoteinbib]{revtex4-2}
\usepackage{tabu,bm}
\usepackage{hyperref}
\usepackage{physics}
\usepackage{graphicx}
\usepackage{capt-of}
\usepackage{enumitem,cleveref}
\DeclareGraphicsExtensions{.pdf,.jpeg,.jpg, .png, .eps, .tiff}
\usepackage{centernot}
\usepackage{multirow}
\usepackage{dsfont}

\usepackage[dvipsnames,table,x11names]{xcolor}
\usepackage{amssymb}

\usepackage[caption=false]{subfig}

\newcommand{\Fuks}{Fuk\'s}
\newcommand{\Fates}{Fat\`es}

\renewcommand{\epsilon}{\varepsilon}

\newcommand{\rarrow}{\rightarrow}

\newcommand{\tx}{\text}

\newcommand{\nn}{\nonumber}

\newcommand{\rom}[1]{\uppercase\expandafter{\romannumeral #1\relax}}

\def\ba#1\ea{\begin{align}#1\end{align}}
\def\bs{\begin{subequations}}
\def\es{\end{subequations}}
\def\beq{\begin{equation}}
\def\eeq{\end{equation}}
\def\bfig{\begin{figure}[h]}
\def\efig{\end{figure}}
\def\bsfig{\begin{subfigure}[b]}
\def\esfig{\end{subfigure}}
\def\bi{\begin{itemize}}
\def\ei{\end{itemize}}

\usepackage{optidef}
\hypersetup{
	colorlinks = true,
	urlcolor = {blue},
	citecolor = {blue},
	linkcolor= {blue}
}
\newcommand{\gkb}[1]{{\textcolor{blue}{#1}}}
\newcommand{\ew}[1]{{\textcolor{magenta}{#1}}}

\begin{document}
\title{Density Classification with Non-Unitary Quantum Cellular Automata}
\author{Elisabeth Wagner}
\email{elisabeth.wagner@students.mq.edu.au}
\affiliation{School of Mathematical and Physical Sciences, Macquarie University, NSW 2109, Australia}
\affiliation{Australian Research Council Centre of Excellence in Engineered Quantum Systems, Macquarie University, Sydney, NSW 2109, Australia}
\author{Federico Dell'Anna}
\email{federico.dellanna2@unibo.it}
\affiliation{Dipartimento di Fisica e Astronomia dell’Universit\`a di Bologna, I-40127 Bologna, Italy}
\affiliation{Istituto Nazionale di Fisica Nucleare, Sezione di Bologna, I-40127 Bologna, Italy}
\author{Ramil Nigmatullin}
\affiliation{School of Mathematical and Physical Sciences, Macquarie University, NSW 2109, Australia}
\affiliation{Quantinuum, 13-15 Hills Road, CB2 1NL Cambridge, United Kingdom}
\author{Gavin K.~Brennen}
\affiliation{School of Mathematical and Physical Sciences, Macquarie University, NSW 2109, Australia}
\affiliation{Australian Research Council Centre of Excellence in Engineered Quantum Systems, Macquarie University, Sydney, NSW 2109, Australia}

\date{\today}

\begin{abstract}

The density classification (DC) task, a computation which maps global density information to local density, is studied using one-dimensional non-unitary quantum cellular automata (QCAs). Two approaches are considered:~one that preserves the number density and one that performs majority voting. For number-preserving DC, two QCAs are introduced that reach the fixed-point solution in a time scaling quadratically with the system size. One of the QCAs is based on a known classical probabilistic cellular automaton which has been studied in the context of DC. 
The second is a new quantum model that is designed to demonstrate additional quantum features and is restricted to only two-body interactions. Both can be generated by continuous-time Lindblad dynamics. A third QCA is a hybrid rule defined by both discrete-time and continuous-time three-body interactions that is shown to solve the majority voting problem within a time that scales linearly with the system size.
\end{abstract}

\maketitle

\onecolumngrid

\section{Introduction}
Cellular automata (CAs) are dynamical systems involving a lattice discretization of space whose multistate cells are updated synchronously based on their own state and the state of the cells within a given radius. For CAs, the dynamics is translationally and temporally invariant, and locality as well as causality are preserved. A benchmark problem for CAs is the density classification (DC) task \cite{fuks,densclassreview2022,laboudi2019,laboudi2020}, which involves mapping the global density of $1$'s of an arbitrary initial configuration of two-state $0/1$ lattice cells to local density information. 

The first (imperfect) density classifier was given by the Gacs--Kurdymov--Levin (GKL) rule \cite{gacs1978}, which has been shown to solve the DC problem within a certain error threshold. When the initial density is close to $0.5$, approximately 70\% of the initial configurations are correctly classified. Attempts to evolve CA that perform the DC task have led to comparable proficiency, classifying correctly about 80\% of all possible initial configurations \cite{mitchell1993arxiv}.

It has been proven \cite{land1995} that there is no one-dimensional two-state, radius $r\geq1$, deterministic CA with periodic boundary conditions that can classify the density of all initial configurations. 
The proof has been generalized by extending the dynamics to both deterministic and probabilistic CAs and to any dimension \cite{busic2013}. By relaxing the assumptions, namely, by adding boundary conditions \cite{sipper1998} or accepting broken translational invariance of the output \cite{capcarrere1996}, a two-state CA can be shown to exist that performs DC with a convergence time $\tau_\tx{conv}$, i.e.,~the number of updates, scaling linearly with the system size $N$.
Moreover, a sequence of two elementary CA rules has been investigated by \Fuks\ that applies first the traffic rule 184 for half the updates and then the majority rule 232 for the remaining steps (hereby, the CA rule 184 eliminates all blocks 00 if the density is greater than $\frac{1}{2}$, or all blocks~11 if less than $\frac{1}{2}$; afterwards, rule 232 yields a homogeneous configuration of all 0~s and all 1~s), 
 which also solves the DC task perfectly with convergence time $\tau_\tx{conv}=N$ \cite{fuks1997}. Later, an investigation of a subset of ternary (three-state) CA rules possessing additive invariants revealed that no absolute DC is possible with a pair of ternary rules \cite{fuks2019}.

If restricting to updates with a single rule, introducing randomness can help. \Fuks\ \cite{fuks} provided a probabilistic CA where the local update rule is non-deterministic that solves the DC task with $\tau_\tx{conv}=\mathcal{O}(N^2)$.
Additionally, \Fates\ \cite{fates} has demonstrated how a stochastic mixture of two deterministic rules, the traffic and majority rules, that is different from the one described in \cite{fuks1997}, can achieve a classification accuracy exceeding 90\%, exhibiting an experimentally confirmed (quasi-)linear scaling of $\tau_\tx{conv}=\mathcal{O}(N)$.
A comprehensive review of the DC problem is given in \cite{DBLP:journals/jca/WolzO08},
and see \cite{DBLP:journals/jca/Oliveira14} for state-of-the-art solutions using cellular~automata.  

Given these CA models established for addressing the DC task, it prompts a natural inquiry into the feasibility and potential efficiency of developing quantum versions of these CA models to serve as density classifiers. This will be especially important if the intent is to use CA-type dynamics on qubits, where the underlying transition rules must obey the laws of quantum mechanics. A straightforward translation of CA update rules to QCAs is not always possible since the former usually involve synchronous updates on all cells, while in the latter, synchronous updates are not allowed and instead one must perform some form of a partitioning of the rule \cite{Schumacher2004ReversibleQC,BW04,costa2018quantum}. Indeed, directly translating local synchronous CA rules to partitioned QCA rules can lead to very different dynamics \cite{PhysRevLett.124.070503}.

Some work on QCAs for DC has been completed recently. Guedes et al.~\cite{guedes2023quantum} introduced two QCAs based on the known elementary CA rule 232 with density-classification capabilities, namely, the local majority voting and the two-line voting. The latter extends rule 232 with an additional temporal dimension. While not a perfect classifier, it was shown to be useful as a way to efficiently perform measurement-free quantum error correction (MFQEC) for bit-flip channels. Their construction can be implemented using local gates in a quasi-1D lattice. In our construction below, only a single 1D lattice is used.

The majority voting problem is closely related to DC, where instead of mapping a global density to a real-valued local density, the majority is mapped to a binary-valued local density. It is a more widely studied task in mathematics and computer science beyond the study of CAs, and has recently been investigated in a range of different contexts using quantum computing algorithms---defining, for example, a quantum-accelerated voting \mbox{algorithm \cite{qmv_liu_2023},} quantum logical veto and nomination rules \cite{qmv_sun_2023}, a quantum parliament~\cite{qmv_adronikus_2022}, a quantum voting protocol which can select multiple winners from candidates \cite{qmv_li_2021}, a non-oracular quantum adaptive search method \cite{zhong2021best}, a quantum majority vote that violates the quantum Arrow's impossibility theorem \cite{qmv_bao_2017}, and a generalized quantum version of the majority vote that determines the majority state given a sequence of quantum states \cite{qmv_buhrman_2022}.

In this work, three QCAs are introduced for these problems. Two of them solve the DC task, where one of them is inspired by the aforementioned CA model by \Fuks\ from 2002 \cite{fuks} and the other one is a new quantum model demonstrating additional quantum features like quantum coherences and correlations in the system and which is restricted to only two-body interactions.
A third QCA is introduced that is designed to address the majority voting problem and is constructed as a hybrid rule. 
Both discrete-time completely positive trace-preserving (CPTP) maps as well as corresponding continuous-time Lindblad dynamics are considered. The efficiency of the first two QCAs in solving the DC task is shown by computing the spectral gap of their respective Lindbladians, while the convergence time of the third QCA is proven for the discrete case.

After presenting the QCA models in Section~\ref{sec:model}, their dynamics are investigated in Section~\ref{sec:results}, before concluding in Section~\ref{sec:conclusion}.

\section{Model} \label{sec:model}
\begin{figure}[h]
    \includegraphics[width=.8\columnwidth]{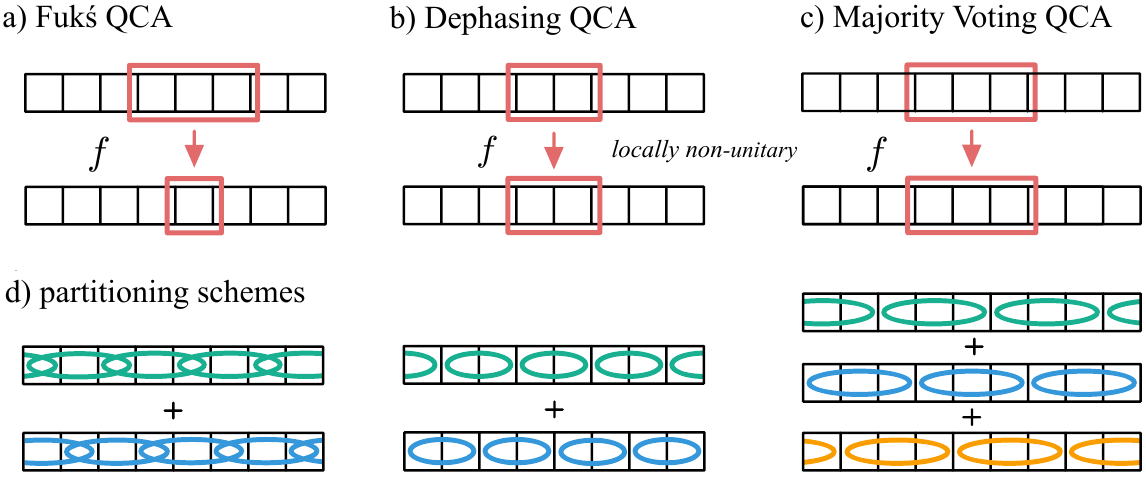}
    \caption{Illustration of the dynamics of (\textbf{a}) the \Fuks\ QCA, (\textbf{b}) the Dephasing QCA, (\textbf{c}) the Majority Voting QCA, and (\textbf{d}) their partitioning schemes with periodic boundaries, where $f$ represents the respective local transition function. While the \Fuks\ QCA is defined by three-body operations where only the center site is updated, all cells of the two-body  neighborhoods are updated for the Dephasing QCA, and likewise all three cells are updated for the Majority Voting QCA. (\textbf{d}) For the \Fuks\ QCA, the three-body operations are applied subsequently onto all even and then all odd lattice sites of the system, whereas in case of the Dephasing QCA and the Majority Voting QCA, only all neighboring non-overlapping neighborhoods can be updated simultaneously. } 
    \label{fig:updates}
\end{figure}

Three non-unitary QCAs are proposed, two of which are density classifiers that conserve the number density of the system and one of which outputs the string with all bits carrying the majority of the input string. The first QCA is inspired by a CA that has been shown to solve the DC task, namely, the ``\Fuks\ CA'' \cite{fuks}, that will be used as a framework to construct a corresponding quantum model; see Section~\ref{sec:fuks}. The second is a novel QCA, called ``Dephasing QCA'', that outperforms the \Fuks\ QCA by only including two-cell interactions; see Section~\ref{sec:dephasing}.
The third QCA is introduced for solving the majority voting problem and is a hybrid rule defined by discrete-time three-body interactions; see \mbox{Section~\ref{sec:mv}}.
All QCAs are defined on a one-dimensional lattice with $N$ lattice sites and periodic boundary conditions; see Figure~\ref{fig:updates}.
To establish the foundational mathematical framework on which this paper is based, the description of the quantum channels is outlined in the following prelude. To start, a quantum channel $\hat{S}$ is in the Kraus decomposition given by 
\ba
    \hat{S}[\hat\rho] = \sum_\mu \hat{K}_{\mu} \, \hat\rho \, \qty(\hat{K}_\mu)^{\!\dagger},
    \label{eq:kraus}  
\ea
where $\big\{\hat{K}_{\mu}\big\}$ labels the set of Kraus operators satisfying the trace-preserving condition
$\sum_\mu \qty(\hat{K}_{\mu})^{\!\dagger}\hat{K}_\mu=\hat{\mathds{1}}$, where $\hat{\mathds{1}}$ is the identity operator.
The quantum density matrices are vectorized using the Choi-isomorphism $\dyad{a}{b}\rarrow\ket{a}\otimes\ket{b}$ such that a density matrix
$\hat{\rho}(t) = \sum_{a,b} \hat{\rho}_{a,b}(t)\dyad{a}{b}$ becomes a vector in a doubled space $\sum_{a,b} \hat{\rho}_{a,b}(t)\ket{a}\otimes\ket{b}$, where the states at each individual site are vectorized first before the tensor product over all sites is applied \cite{gilchrist11}. Under this mapping, the Kraus decomposition \eqref{eq:kraus} becomes
\ba
    \mathbb{\hat{S}} = \sum_\mu \hat{K}_{\mu} \otimes \qty(\hat{K}_{\mu})^{\!*} ,
    \label{eq:krausVec}  
\ea
which acts on the doubled Hilbert space $\mathcal{H}=\prod_{j}\mathcal{H}_j\otimes\mathcal{H}_j^*$, where $\mathcal{H}_j^*$ denotes the dual Hilbert space on site $j$.
Furthermore, for considering continuous-time dynamics, the Lindblad evolution
\ba
    \hat{\mathcal{L}}[\hat\rho]
    =
    -i\qty[\hat H,\hat\rho]
    +
    \sum_{j=1}^{N} \sum_k
    \qty(\hat L_{k_j} \hat\rho\hat L_{k_j}^\dagger - \frac{1}{2} \qty(\hat L_{k_j}^\dagger \hat L_{k_j} \hat\rho + \hat\rho \hat L_{k_j}^\dagger \hat L_{k_j} ))
\label{eq:Lio}
\ea
is utilized, where $i$ labels the imaginary unit, $\hat H$ represents the Hamiltonian, and $\big\{\hat{L}_{k_j}\big\}$ is the set of jump operators acting on lattice site $j$ (henceforth, we set $\hbar\equiv 1$). In the vectorized form, Equation~\eqref{eq:Lio} becomes
\ba
 \hat{\mathbb{L}}= 
 -i\qty(\hat H \otimes \hat{\mathds{1}}-\hat{\mathds{1}}\otimes \hat H^T) + 
 \sum_{j=1}^{N}\sum_k \qty(  \hat L_{k_j} \otimes \hat L_{k_j}^*
 -\frac{1}{2} \qty(\hat L_{k_j}^{\dagger}\hat L_{k_j} \otimes  \hat{\mathds{1}}+  \hat{\mathds{1}} \otimes \hat L_{k_j}^{\dagger}\hat L_{k_j})).
\label{eq:LioVec}
\ea

Given this framework, the theoretical description of the proposed QCA models is outlined next.

\subsection{\Fuks\ QCA} \label{sec:fuks}

The \Fuks\ rule \cite{fuks} is a radius-one probabilistic CA given by the transition probabilities presented in Table~\ref{tab:fuks}.
A cell in state one with two neighboring zero states becomes zero with probability $2p$, and, analogously, a cell in the zero state surrounded by two one states becomes a one state with the same probability $2p$. If the neighboring sites are in two different states, then the state at the center site is flipped with probability $p$.
Zero (one) states are mapped to one (zero) states with a probability proportional to the number of ones (zeroes) in the neighborhood. It was shown that the dynamics of the local density can be approximated by the standard diffusion equation, implying that the convergence time scales quadratically with the system size $N$, $\tau_\tx{conv}=\mathcal{O}(N^2)$. Deriving quantum dynamics inspired by the \textit{{non-partitioned}} \Fuks\ CA faces the challenge that CAs are implemented by making a copy of the whole state at each time step---because only then could all cells, at both the even and odd lattice sites, be updated simultaneously based on the neighboring states at the previous time step. This copy operation, as a fundamental part of classical CAs, can, however, not be performed on a quantum state due to the no-cloning \mbox{theorem \cite{nocloning}.} 
Only specifically \textit{{partitioned}} CAs that update all even and all odd sites one after the other in consecutive time steps and do not involve the implementation of a copy process can be directly translated into a corresponding QCA. An example of a partitioned CA is the Domany--Kinzel CA model, originally proposed in \cite{DomanyKinzel}, whose associate quantum version has been intensively investigated in i.a.
~\cite{lesanovsky_2019, Gillman2020,Gillman2021, Gillman2021x,DP}. On the other hand, CAs that are not partitioned (and do include the copy process) could not be directly translated into a quantum map. 
This is why the definitions of the quantum channel are in this work merely inspired by the basic framework of the considered non-partitioned CAs.

\begin{table}[h]
\begingroup
\setlength{\tabcolsep}{10pt} 
\renewcommand{\arraystretch}{1.5} 
\begin{center}
	\begin{tabular}{c|c|l|c}
    neighborhood & transition & probability & Kraus operators \\\hline\hline
        \multirow{2}{*}{00} &
        $000 \rarrow 010$ &
        $p_{000} = 0$ & amplitude damping \\
        &
        $010 \rarrow 010$ &
        $p_{010} = 1-2p$ & $\Big\{\hat P_0+\sqrt{1-2p}\,\hat P_1,\sqrt{2p}\,\hat\sigma^-\Big\}$ \\\hline
        \multirow{2}{*}{01} &
        $001 \rarrow 011$ &
        $p_{001} = p$ & \\
        &
        $011 \rarrow 011$ &
        $p_{011} = 1-p$ & stochastic bit-flip \\\cline{1-3}
        \multirow{2}{*}{10} &
        $100 \rarrow 110$ &
        $p_{100} = p$ & $\Big\{\sqrt{1-p}\,\hat{\mathds{1}},\sqrt{p}\,\hat{X}\Big\}$ \\
        &
        $110 \rarrow 110$ &
        $p_{110} = 1-p$ & \\\hline
        \multirow{2}{*}{11} &
        $101 \rarrow 111$ & $p_{101} = 2p$ & amplitude pumping \\
        & $111 \rarrow 111$ &
        $p_{111} = 1$ & $\Big\{\sqrt{1-2p}\,\hat P_0+ \hat P_1,\sqrt{2p}\,\hat\sigma^+\Big\}$
	\end{tabular}
    \end{center}
	\caption{\Fuks\ QCA. 
 The transition probabilities $p_{acb}$ represent the likelihood of the state transition $\ket{acb}\rightarrow\ket{a1b}$, with $a,b,c\in\{0,1\}$ $\forall \ p\in\big(0,\frac{1}{2}\big]$. Note that the associate input/output states are two-on-one with the output center site set to be in the one state; the transition from the same input state to the corresponding output state with the center site in the zero state is, correspondingly, one minus the associate transition probability (for example, the transition $110\rarrow 100$ occurs with probability $1-p_{110}=1-(1-p)=p$). Fourth column: set of Kraus operators of the associated quantum channels acting on the center site $j$, where $\hat P_0=\dyad{0}$, $\hat P_1=\dyad{1}$, $\hat{\mathds{1}}=\dyad{0}+\dyad{1}$, $\hat\sigma^-=\dyad{0}{1}$, and $\hat\sigma^+=\dyad{1}{0}$.}
\label{tab:fuks}
\endgroup
\end{table}

A quantum version based on the \Fuks\ CA is defined by
\ba
    \mathbb{\hat S}^\tx{(\Fuks)}
    =\prod_j \qty( 
    \mathbb{\hat S}_j^{(00)} +
    \mathbb{\hat S}_j^{(01)} +
    \mathbb{\hat S}_j^{(10)} +
    \mathbb{\hat S}_j^{(11)} ),
\label{eq:s}
\ea
where
\ba
    \mathbb{\hat S}_j^{(ab)} = 
    \dyad{aa}_{j-1} \otimes 
    \mathbb{\hat{K}}^{(ab)}
    \otimes \dyad{bb}_{j+1},
    \label{eq:s2}
\ea
and
\ba
    \mathbb{\hat{K}}^{(ab)} = \sum_\mu \hat{K}_{\mu}^{(ab)} \otimes \qty(\hat{K}_{\mu}^{(ab)})^{\!*} \ \ \forall\ a,b\in\{0,1\}.
\ea

Note that each local operator of the superoperator \eqref{eq:s} acts non-trivially only on the three-cell neighborhood ($j-1,j,j+1$) of the lattice, thereby preserving locality as well as spatial and temporal invariance by performing the same operation on all sites during each QCA update. The projectors $\dyad{aa}$ and $\dyad{bb}$ that act on the left ($j-1$) and the right ($j+1$) sites determine the neighborhood of the qubit at the center site $j$, on which the superoperator acts. The definition of the four sets of Kraus operators $\big\{\hat{K}_{\mu}^{(ab)}\big\}$ thus fully defines the quantum channel. For the \Fuks\ QCA, these are given by 
\bs
\ba
    \hat K_{0}^\tx{(00)}
        &= \hat P_0+\sqrt{1-2p}\,\hat P_1, 
    &&\hat K_{1}^\tx{(00)}
        = \sqrt{2p} \ \hat\sigma^-, \label{eq:krausfuks00}\\
    \hat K_{0}^\tx{(01)}
        &=  \sqrt{1-p}\ \hat{\mathds{1}}, 
    &&\hat K_{1}^\tx{(01)}
        =  \sqrt{p}\ \hat{X},\label{eq:krausfuks01} \\
    \hat K_{0}^\tx{(10)}
        &=  \sqrt{1-p}\ \hat{\mathds{1}}, 
    &&\hat K_{1}^\tx{(10)}
        =  \sqrt{p}\ \hat{X}, \\
    \hat K_{0}^\tx{(11)}
        &= \hat P_1+\sqrt{1-2p}\,\hat P_0,
    &&\hat K_{1}^\tx{(11)}
        = \sqrt{2p} \ \hat\sigma^+, \label{eq:krausfuks4}
\ea
\label{eq:krausfuks}\es
which satisfy the trace-preserving condition $\sum_{\mu=0,1} \qty(\hat{K}_{\mu}^{(ab)})^{\!\dagger}\hat{K}_{\mu}^{(ab)}=\hat{\mathds{1}}$, where $p\in\big(0,\frac{1}{2}\big]$, $\hat P_0=\dyad{0}$, $\hat P_1=\dyad{1}$ $\hat{\sigma}^-=\dyad{0}{1}$, $\hat{\sigma}^+=\dyad{1}{0}$, and $\hat X$ is the Pauli-X operator. Another way to describe the CA is by applying, for each cell $j$ independently, the elementary CA rule 170 with probability $p$, rule 240 with the same probability $p$, and the identity operation with probability $1-2p$; see p.~230 in \cite{fates}:
\ba
    \mathbb{\hat S}^\tx{(\Fuks)}
    =
    p \ \mathbb{\hat S}^{(170)}
    + p \ \mathbb{\hat S}^{(240)}
    + (1-2p) \ \mathds{\hat 1}.
\label{eq:cafuks}
\ea

Associated continuous-time dynamics are described by the Lindbladian
\ba
    \hat{\mathbb{L}}^\tx{(\Fuks)}
    =
    \sum_{j=1}^{N} \sum_{k=1}^{6}\qty(  \hat L_{k_j} \otimes \hat L_{k_j}^*
     -\frac{1}{2} \qty(\hat L_{k_j}^{\dagger}\hat L_{k_j} \otimes  \hat{\mathds{1}}+  \hat{\mathds{1}} \otimes \hat L_{k_j}^{\dagger}\hat L_{k_j}) ),
\label{eq:LioFuks}
\ea
with the six jump operators
\bs
\ba
   \hat{L}_{1_j} &= \sqrt{\gamma} \dyad{0}_{j-1} \otimes \hat{\sigma}^-_j \otimes \dyad{0}_{j+1}, \label{eq:Fuksjumpop1}\\
   \hat{L}_{2_j} &= \sqrt{\frac{\gamma}{2}}\dyad{0}_{j-1} \otimes \hat{\sigma}^-_j \otimes \dyad{1}_{j+1},\label{eq:Fuksjumpop2} \\
   \hat L_{3_j} &= \sqrt{\frac{\gamma}{2}}\dyad{0}_{j-1} \otimes \hat{\sigma}^+_j \otimes \dyad{1}_{j+1},\label{eq:Fuksjumpop3} \\
   \hat L_{4_j} &= \sqrt{\frac{\gamma}{2}}\dyad{1}_{j-1} \otimes \hat{\sigma}^-_j \otimes \dyad{0}_{j+1}, \\
   \hat L_{5_j} &= \sqrt{\frac{\gamma}{2}}\dyad{1}_{j-1} \otimes \hat{\sigma}^+_j \otimes \dyad{0}_{j+1}, \\
   \hat L_{6_j} &= \sqrt{\gamma} \dyad{1}_{j-1} \otimes \hat{\sigma}^+_j \otimes \dyad{1}_{j+1}.\label{eq:Fuksjumpop6}
\ea
\label{eq:Fuksjumpops}\es \vspace{-12pt}

The jump operators $\hat{L}_{1_j}$ and  $\hat L_{6_j}$ ensure the amplitude damping/pumping transitions $010\rightarrow000$ and $101\rightarrow111$ for long-time evolution $\tau\gg1/\gamma$. The other jump operators $\hat{L}_{2_j}$ to $\hat L_{5_j}$ simulate the bit-flip channel in case of the 01 and 10 neighborhoods, where the overall scaling factor $\frac{1}{\sqrt{2}}$ ensures that the bit-flip operation is implemented with half the probability compared to the amplitude damping/pumping operations---this is analogous to the classical \Fuks\ CA that implements the bit-flip with probability $p$ and the amplitude damping/pumping with probability $2p$; see Table~\ref{tab:fuks} and the derivation in Appendix~\ref{app:decayrate}. 
Note that by setting the decay rate $\gamma\equiv1$ in all calculations, the convergence time of the system can thus be determined as multiples of the time steps $\tau$.
Furthermore, as illustrated on the left in Figure~\ref{fig:updates}d, the \Fuks\ QCA is approximated by a partitioning scheme which is enhanced by repeatedly updating all even and then all odd lattice sites with infinitesimal time updates $\tau$:
\ba
    e^{\hat{\mathbb{L}}^\tx{(even)}\tau}
    e^{\hat{\mathbb{L}}^\tx{(odd)}\tau}
    \approx
    e^{\qty(\hat{\mathbb{L}}^\tx{(even)}+\hat{\mathbb{L}}^\tx{(even)})\tau},
    \label{eq:Levenodd}
\ea
where $\hat{\mathbb{L}}^\tx{(even/odd)}$ describes the \Fuks\ Lindbladian \eqref{eq:LioFuks} acting on all even/odd lattice sites simultaneously.
\subsection{Dephasing QCA} \label{sec:dephasing}
While the \Fuks\ QCA is inspired by a classical CA, the here-introduced quantum model, dubbed the Dephasing QCA, is more efficiently constructed since the local map requires only two-body interactions. This rule preserves the number density of the input state and maps the system's global number density to the local density information.

The Dephasing QCA is given by the Lindblad evolution
\begin{equation}
\begin{split}
    \hat{\mathbb L}^\tx{(Dephasing)} =&
    -i\qty(\hat H \otimes \hat{\mathds{1}}-\hat{\mathds{1}}\otimes \hat H^T)
     \\ &\ \  \ +
    \sum_{j=1}^N \sum_{k=1}^4
    \qty(  \hat L_{k_{j,j+1}} \otimes \hat L_{k_{j,j+1}}^*
    -\frac{1}{2} \qty(\hat L_{k_{j,j+1}} ^{\dagger}\hat L_{k_{j,j+1}}  \otimes  \hat{\mathds{1}}+  \hat{\mathds{1}} \otimes \hat L_{k_{j,j+1}} ^{\dagger}\hat L_{k_{j,j+1}} ) ) 
    \label{eq:LioDephasing}
\end{split}
\end{equation}
with Hamiltonian
\begin{equation}
    \hat H = \Omega\sum_{j=1}^N \qty(\hat X_{j}\hat X_{j+1}+\hat Y_{j}\hat Y_{j+1}),\label{eq:HDephasing}    
\end{equation}
where $\Omega\in\mathbb{R}$, $\hat X$ and $\hat Y$ represent the associated Pauli operators, and the jump operators  $\hat{L}_{k_{j,j+1}}$ act each on the two neighboring sites, $j$ and $j+1$, where $j+1\equiv1$ if $j=N$ considering periodic boundary conditions. The latter are given by the four projectors 
\bs
\ba
   \hat L_{1_{j,j+1}} &= \dyad{00}_{j,j+1}, \label{eq:jumpopsDephasing1}\\
   \hat L_{2_{j,j+1}} &= \dyad{\psi^+}_{j,j+1}, \label{eq:jumpopsDephasing2}\\
   \hat L_{3_{j,j+1}} &= \dyad{\psi^-}_{j,j+1}, \label{eq:jumpopsDephasing3}\\
   \hat L_{4_{j,j+1}} &= \dyad{11}_{j,j+1},\label{eq:jumpopsDephasing4}
\ea
\label{eq:jumpopsDephasing}\es
with the Bell states $\ket{\psi^\pm}=\frac{1}{\sqrt{2}}(\ket{01}\pm\ket{10})$. 
Note that the QCA acts in the same way on the left and on the right site of each two-cell neighborhood and that the dissipator is parity symmetric.
The projectors are eigenstates of the Hamiltonian and are designed to remove coherences between different eigenspaces of $\hat S_z$, but also within the same eigenspace of $\hat S_z$.

The corresponding partitioning scheme of the QCA is illustrated and described in Figure~\ref{fig:updates}d, where the two sets of two-body updates are, analogous to the \Fuks\ QCA, approximated by infinitesimal time updates generated by Lindbladians according to \mbox{Equation~\eqref{eq:Levenodd}}, where the even (odd) updates are here defined to be those where the left cells of the two-body neighborhoods are located at the even (odd) lattice sites, and the neighborhoods do not overlap in one partial time step.

\subsection{Majority Voting QCA} \label{sec:mv}
For the task of majority voting, analogously to what was performed for the \Fuks\ rule above, one might try to use a quantum version of the 
 \Fates\ CA \cite{fates} rule. However, a direct construction does not work as described in Appendix~\ref{sec:fates}.\\
Therefore, a new solution is proposed. This solution requires relaxing the strict definition of CA, in which only the central cell is updated. Furthermore, since our goal is to classify the initial state based on whether its initial density is greater or less than $N/2$, the idea is to structure what differentiates these two sectors.
Let $n$ be the expectation value of $\hat{n}=\sum_j \dyad{1}_j$. It is easy to observe that if $n \leq N/2$, it will always be possible to distribute the ones along the chain in such a way as to avoid them being neighboring. For $n > N/2$, this is no longer possible. The idea is to define a transformation  $\hat{\mathbb{A}}$ such that the following~hold:
\begin{itemize}
	\item Its repeated action on a state $\ket{\hat\rho}$ spreads the $\ket{1}$ states out along the chain so that the final state obtained does not exhibit two neighboring $\ket{1}$ states.
	 \item It satisfies $\qty[\hat{\mathbb{S}}_z, \hat{\mathbb{A}}]=0$, where $\hat{\mathbb{S}}_z$ is the vectorized form of $\hat{S}_z=\frac{1}{2}\sum_j \hat Z_j$ with Pauli operator $\hat Z_j$, which will preserve the number density in the system.
\end{itemize} 

Thus, we define this transformation as
\begin{equation}
    \hat{\mathbb{A}} =\prod_j \sum_{\mu=0,1} \qty(\hat{K}_{\mu_j}\otimes \hat{K}_{\mu_j}^{*} ),
 \label{eq:A}
\end{equation}
where
\bs
\ba
	 \hat{K}_{0_j} &=\dyad{1}_{j-1} \otimes \dyad{0}{1}_j \otimes \dyad{1}{0}_{j+1}, \\
	\hat{K}_{1_j} &= \hat{\mathds{1}} - \dyad{1}_{j-1} \otimes \dyad{1}_j\otimes \dyad{0}_{j+1}
\ea\es
satisfy the trace-preserving condition $\sum_{\mu=0,1} \hat{K}_{\mu}^\dagger \;\hat{K}_{\mu} =\hat{\mathds{1}}$. The proof for $\hat{\mathbb{A}}$ satisfying the two aforementioned properties can be found in Appendix~\ref{app:MV_A}.
In Equation~\eqref{eq:A}, each factor in the product does not commute with its nearest neighbors nor with its next-to-nearest neighbors but rather with every third site. This implies that different orders of these factors lead to different versions of $\hat{\mathbb{A}}$. However, each of them satisfies the aforementioned requirements such that it is convenient to choose the one that maximizes the number of operations in a single time step:
\begin{equation}
	\hat{\mathbb{A}} \longrightarrow
	\hat{\mathbb{A}}^\text{(1)}
	\hat{\mathbb{A}}^\text{(2)}
	\hat{\mathbb{A}}^\text{(3)} 
	\label{eq:A123}
\end{equation}
where $\hat{\mathbb{A}}^{(x)}$ with $x\in\{1,2,3\}$ describes the action on the associate sets of neighboring, non-overlapping three-cell neighborhoods; see the illustration in Figure~\ref{fig:updates}d.

Once transformation \( \hat{\mathbb{A}} \) is applied, the resulting state must be brought to \( |0\rangle^{\otimes N} \) if it does not contain any cluster of \( |1\rangle \)s. Otherwise, such a cluster must be progressively expanded until it covers the entire chain and reaches the state \( |1\rangle^{\otimes N} \). This can be obtained by applying repeatedly
\begin{equation}
	\hat{\mathbb{B}} =\prod_j \sum_{\mu=0,1,2,3} \qty(\hat{K}_{\mu_j}\otimes \hat{K}_{\mu_j}^{*}),
\end{equation}
where
\bs
\ba
\hat{K}_{0_j} &=\dyad{0}_{j-1} \otimes \dyad{0}{1}_j \otimes \dyad{0}_{j+1}, \\
\hat{K}_{1_j} &=\dyad{1}_{j-1} \otimes \dyad{1}_j \otimes \dyad{1}{0}_{j+1}, \\
\hat{K}_{2_j} &=\dyad{1}{0}_{j-1} \otimes \dyad{1}_j \otimes \dyad{1}_{j+1}, \\
\hat{K}_{3_j} &= \hat{\mathds{1}} -( \dyad{0}_{j-1} \otimes \dyad{1}_j\otimes \dyad{0}_{j+1} + \dyad{1}_{j-1} \otimes \dyad{1}_j\otimes \dyad{0}_{j+1} \\ & \quad +\dyad{0}_{j-1} \otimes \dyad{1}_j\otimes \dyad{1}_{j+1}), \nonumber
\ea
\es
that satisfy the trace-preserving condition $\sum_{\mu=0,1,2,3} \hat{K}_{\mu}^\dagger \;\hat{K}_{\mu} =\hat{\mathds{1}}$. Similar to $\hat{\mathbb{A}}$, we adopt a non-overlapping three-cell partition pattern for $\hat{\mathbb{B}}$.
The minimum number of times that  $\hat{\mathbb{A}}$ and  $\hat{\mathbb{B}}$ need to be applied ($m_a$ and $m_b$) depends on the specific partition scheme chosen as well as the initial state. In Appendix~\ref{app:proofAB}, we derive the minimum number of layers with respect to our partition scheme capable of classifying every initial state. \\In summary, our proposal to solve the majority voting problem is
\begin{equation}
\begin{array}{cl}
\hat{\mathbb{B}}^{m_b} \hat{\mathbb{A}}^{m_a} \ket{\hat\rho}&= 
\underbrace{\hat{\mathbb{B}}^{(1)}
	\hat{\mathbb{B}}^{(2)}
	\hat{\mathbb{B}}^{(3)}}_{m_b} \
\underbrace{\hat{\mathbb{B}}^{(1)}
	\hat{\mathbb{B}}^{(2)}
	\hat{\mathbb{B}}^{(3)}}_{m_b-1} 
\	\cdots \
\underbrace{\hat{\mathbb{B}}^{(1)}
	\hat{\mathbb{B}}^{(2)}
	\hat{\mathbb{B}}^{(3)}}_{2} \
\underbrace{\hat{\mathbb{B}}^{(1)}
	\hat{\mathbb{B}}^{(2)}
	\hat{\mathbb{B}}^{(3)}}_{1}   \\
	& \quad
\times\underbrace{\hat{\mathbb{A}}^{(1)}
	\hat{\mathbb{A}}^{(2)}
	\hat{\mathbb{A}}^{(3)}}_{m_a} \
\underbrace{\hat{\mathbb{A}}^{(1)}
	\hat{\mathbb{A}}^{(2)}
	\hat{\mathbb{A}}^{(3)}}_{m_a-1} 
\ \cdots \
\underbrace{\hat{\mathbb{A}}^{(1)}
	\hat{\mathbb{A}}^{(2)}
	\hat{\mathbb{A}}^{(3)}}_{2} \
\underbrace{\hat{\mathbb{A}}^{(1)}
	\hat{\mathbb{A}}^{(2)}
	\hat{\mathbb{A}}^{(3)}}_{1} \ket{\hat\rho}. 
\end{array}
\label{eq:MV_discrete}
\end{equation}

In addition, it is possible to define two Lindbladian operators $\mathcal{L}_A$ and $\mathcal{L}_B$ capable of effecting the continuous-time evolution of transformations $\hat{\mathbb{A}}$ and $\hat{\mathbb{B}}$, which, in the vectorized form, are
\begin{equation}
	\hat{\mathbb{L}}^\text{A}
	=
	\sum_{j=1}^{N} \qty(  \hat L_{0_j}^{a} \otimes \hat L^{a}_{0_j}
	-\frac{1}{2} \qty(\hat L_{0_j}^{a\dagger}\hat L_{0_j}^{a} \otimes  \hat{\mathds{1}}+  \hat{\mathds{1}} \otimes \hat L_{0_j}^{a\dagger}\hat L_{0_j}^{a}) ),
	\label{eq:LA}
\end{equation}
\begin{equation}
    \hat{\mathbb{L}}^\text{B}
=
\sum_{j=1}^{N} \sum_{k=0}^{2}\qty(  \hat L_{k_j}^{b} \otimes \hat L^{b}_{k_j}
-\frac{1}{2} \qty(\hat L_{k_j}^{b\dagger}\hat L_{k_j}^{b} \otimes  \hat{\mathds{1}}+  \hat{\mathds{1}} \otimes \hat L_{k_j}^{b\dagger}\hat L_{k_j}^{b}))
\label{eq:LB}
\end{equation}
where 
\bs \ba
	 \hat{L}^{a}_{0_j} &=\dyad{1}_{j-1} \otimes \dyad{0}{1}_j \otimes \dyad{1}{0}_{j+1} \\
	\hat{L}^{b}_{0_j} &=\dyad{0}_{j-1} \otimes \dyad{0}{1}_j \otimes \dyad{0}_{j+1}, \\
	\hat{L}^{b}_{1_j} &=\dyad{1}_{j-1} \otimes \dyad{1}_j \otimes \dyad{1}{0}_{j+1}, \\
	\hat{L}^{b}_{2_j} &=\dyad{1}{0}_{j-1} \otimes \dyad{1}_j \otimes \dyad{1}_{j+1}.
\ea \es
Then, our proposal to solve the majority voting problem, by using the continuous-time evolution, is
\begin{equation}
e^{\hat{\mathbb{L}}^{\text{B}}\tau_B} e^{\hat{\mathbb{L}}^\text{A}\tau_A} \ket{\hat{\rho}},
\label{eq:continuous_time}
\end{equation}
where $\tau_A$ represents the time needed to reach a state without two adjacent $|1\rangle$s, and $\tau_B$ represents the time to expand a cluster of $|1\rangle$s along the entire chain, both in the worst-case~scenario.

Note the jump operators in Equation~(\ref{eq:LB})  differ from the corresponding jump operators in Equation~(\ref{eq:LioFuks}), in that they do not restrict to projectors on the left and right cells, and hence have discrete evolution that is less parallelizable. An attempt was made to find jump operators like in the \Fuks\ rule here by using a supervised machine learning approach. However, this method yielded only a partial solution with extremely long convergence times, and it was not further explored. Nevertheless, a detailed description of this approach and its results can be found in Appendix~\ref{app:LMV}.

\section{Results} \label{sec:results}

Next, the research results on the three QCA models are presented: the \Fuks\ QCA in Section~\ref{sec:results_fuks}, the Dephasing QCA in Section~\ref{sec:results_dephasing}, and the Majority Voting QCA in Section~\ref{sec:results_mv}.

\subsection{\Fuks\ QCA} \label{sec:results_fuks}
The dynamics of the \Fuks\ QCA are elaborated in the following; see the definition in Section~\ref{sec:fuks}. It is shown that $\hat{\mathbb{L}}^\tx{(\Fuks)}[\hat\rho]$ conserves the number density of the initial state $\hat\rho$ in analogy to the associate classical CA rule.
The number density can be quantified by the operator $\hat{S}_z=\frac{1}{2}\sum_j \hat Z_j$, whose expectation value is conserved as
\ba
    \frac{\tx d}{\tx dt}\expval{\hat{S}_z(t)} = 0,
\ea
see the proof in Appendix~\ref{app:SzFuks}.
Furthermore, the \Fuks\ Lindbladian in Equation~\eqref{eq:LioFuks} exhibits four zero eigenvalues that correspond to the set of steady states 
\ba
    \hat{\rho}_\tx{ss}^\tx{(\Fuks)}
    &=(1-\alpha)\dyad{0...0}
    +\beta\dyad{0...0}{1...1} 
    +\beta^*\dyad{1...1}{0...0}
    +\alpha\dyad{1...1},
    \label{eq:ssfuks}
\ea
where $\alpha\in[0,1]$ represents the global (and local) number density of the state, and $\beta,\beta^*\in\mathbb{C}$ are the amplitudes of the off-diagonal coherence terms; see the proof in Appendix~\ref{app:steadystatesfuks}.
Note that the pure states $\ket{0...0}$ and $\ket{1...1}$, as well as the GHZ state are included in this set corresponding to the parameter sets $\{\alpha=0,\beta=0\}$, $\{\alpha=1,\beta=0\}$, and $\{\alpha=\frac{1}{2},\beta=\frac{1}{2}\}$, respectively. All off-diagonal elements unequal to $\dyad{0...0}{1...1}$ or $\dyad{1...1}{0...0}$ are shown to decohere under the action of this map as derived in Appendix~\ref{app:appss2}.
As an example for the dynamics of this QCA, the initial states $\ket{001}$ and $\ket{011}$ are considered that would, in the long-time limit $t\gg1$, evolve to the following steady states:\vspace{6pt}
\bs
\ba
    \dyad{001}&\rightarrow
    \frac{2}{3}\dyad{000}+\frac{1}{3}\dyad{111},\label{eq:example001}\\
    \dyad{011} &\rightarrow
    \frac{1}{3}\dyad{000}+\frac{2}{3}\dyad{111},
\ea\es
where the global number densities of $\frac{1}{3}$ and $\frac{2}{3}$, respectively, are conserved.

\begin{figure}[h]

\includegraphics[width=0.5\columnwidth]{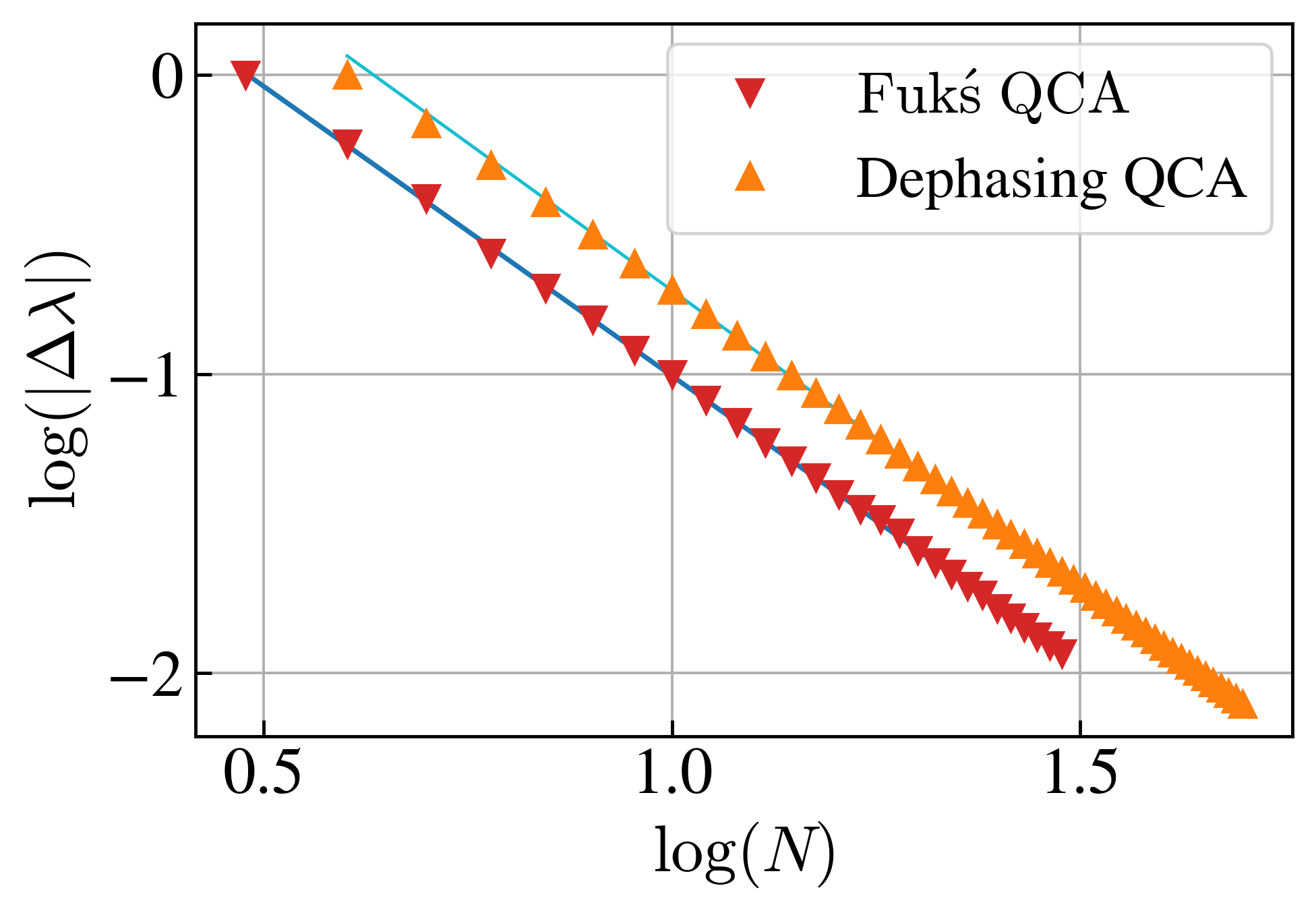}
     \caption{ Logarithmic plot of the spectral gap $\Delta\lambda$ versus the system size $N$ for the Lindbladians $\hat{\mathcal{L}}^\tx{(\Fuks)}$ and $\hat{\mathcal{L}}^\tx{(Dephasing)}$, see Equations~\eqref{eq:LioFuks} and \eqref{eq:LioDephasing}, respectively. For the Dephasing QCA, the Hamiltonian is turned off $(\Omega=0)$. Using the DMRG algorithm \cite{dmrg}, the spectral gap is computed for system sizes $N\in[3,30]$  for the \Fuks\ QCA, and $N\in[4,50]$ for the Dephasing QCA.
     The subjacent blue and cyan lines represent the corresponding linear regression fits $\log(|\Delta\lambda|)=c\times\log(N)+d$ with parameters 
 $c=-1.937\pm5\times 10^{-3}$ and $d=0.931\pm6\times 10^{-3}$ for the \Fuks\ QCA, and $c=-1.972\pm3\times 10^{-3}$ and $d=1.252\pm4\times 10^{-3}$ for the Dephasing QCA.
     For the latter, the first two points of the spectral gap corresponding to $N=4,5$ are excluded from the calculation of the linear regression, which has halved the associate standard deviation of the slope. }
 \label{fig:spectralgap}
 \end{figure}

For quantifying the convergence time $\tau_\tx{conv}$, i.e.,~the maximum time to reach the steady state of the system, the spectral gap $\Delta\lambda$ is determined. The latter is the energy difference between the ground state and the first excited state, and is given by the smallest non-zero absolute value of the eigenvalues of the Lindbladian. Note that all non-zero eigenvalues are negative such that the spectral gap corresponds to the negative of largest non-zero eigenvalue. A logarithmic plot of the spectral gap versus the system size is shown in Figure~\ref{fig:spectralgap}.

 An almost quadratic inverse scaling of the spectral gap with the system size is observed, $\Delta\lambda\propto N^{-1.942\pm0.005}$, such that the convergence time scales almost quadratically with the system size:
\ba
    \tau_\tx{conv}
    \propto\mathcal O\qty(\frac{1}{\Delta\lambda})
    \approx\mathcal O\qty(N^2).
    \label{eq:tauconv}
\ea

\subsection{Dephasing QCA}\label{sec:results_dephasing}

Next, the results of the Dephasing QCA are outlined; see the definition in Section~\ref{sec:dephasing}. It is derived that the number density of the system is conserved with
\ba
    \frac{\tx d}{\tx dt}\expval{\hat{S}_z(t)} = 0,
\ea
and that the Dephasing QCA indeed solves the DC task; see the proof in Appendix~\ref{app:Dephasing}.
To exemplify the dynamics of this QCA, the initial states $\ket{001}$ and $\ket{011}$ are considered that would, in the long-time limit $t\gg1$, evolve to the following mixed steady states:\vspace{6pt}
\bs
\ba
    \dyad{001} &\rightarrow \frac{1}{3}(\dyad{001} + \dyad{010} + \dyad{100}), \\
    \dyad{011} &\rightarrow \frac{1}{3} (\dyad{011} + \dyad{101} + \dyad{110}).
\ea\es

For determining the convergence time $\tau_\tx{conv}$, the spectral gap $\Delta\lambda$ is computed, mirroring the approach taken for the \Fuks\ QCA in the previous subsection. The result is presented in Figure~\ref{fig:spectralgap}, where the slope of the linear regression fit shows that $\Delta\lambda\propto N^{-1.972\pm0.003}$ such that the convergence time $\tau_\tx{conv}$ scales almost quadratically with the system size $N$ similar to the \Fuks\ QCA; see Equation~\eqref{eq:tauconv}. However, the spectral gap is by a constant factor of $0.321\pm9\times 10^{-3}$ larger than the spectral gap of the \Fuks\ QCA, which implies that $\tau_\tx{conv}$ is reduced (i.e.,~improved) by this factor in comparison to the \Fuks\ QCA.
When including the Hamiltonian \eqref{eq:HDephasing}, numerical simulations indicate that the scaling of the convergence time $\tau_\tx{conv}$ with $N$ remains unaltered.
Note that by the nature of the dephasing terms and the Hamiltonian, the projectors will remove coherence terms and lead to a fixed point that is block diagonal in $\hat S_z$. That is, if the Hamiltonian is non-zero, the fixed point will be block diagonal with, potentially, coherence terms in each block.

\subsection{Majority Voting QCA}\label{sec:results_mv}
In the following, the dynamics of the Majority Voting QCA are discussed, see Section~\ref{sec:mv}. Our discrete-time evolution proposal consists of a repeated application of $\hat{\mathbb{A}}$ and, subsequently, $\hat{\mathbb{B}}$ (see  Equation~\eqref{eq:MV_discrete}) with a non-overlapping three-cell partition pattern as shown in Figure~\ref{fig:updates}d, allowing for the correct classification of every initial state. We consider the application of a single layer per unit time, so the time required to reach the final state, in the worst-case scenario and with $N\text{mod}(3)=0$, scales in the following way with the system size:
\begin{equation}
\tau =\tau_A +\tau_B = 4  \left\lfloor \frac{N}{2} \right\rfloor +\frac{2}{3}N - 5.
\label{eq:scaling_discrete}
\end{equation}
The proof of this equation can be found in Appendix~\ref{app:proofAB}.

If $N\text{mod}(3)=1$ (or $2$), the partition scheme will have $1$ (or $2$) non-updated cell(s) at each layer. To prevent the same cells from remaining non-updated each time, one could periodically shift the partition scheme so that these cells change over time, traversing through the chain. However, in these cases, it is challenging to establish the worst-case scenario to provide a sufficient value of $\tau$ valid for all initial states. Additionally, we have observed that starting from certain initial states, delays due to the lack of updating some cells scale linearly with $N$. 
This is sufficient to propose a more efficient solution: if $N \text{mod}(3)=1$, one can simply add two extra qubits (one in $|0\rangle$ and the other in $|1\rangle$) and evolve the entire system; if $N\text{mod}(3)=2$, one can add four extra qubits (two in $|0\rangle$ and two in $|1\rangle$) and evolve the entire system. This approach enables us to achieve systems with $N\text{mod}(3)=0$ without altering the initial majority of $\ket{0}$s or $\ket{1}$s.

\begin{figure}[h]
    \includegraphics[scale=0.4]{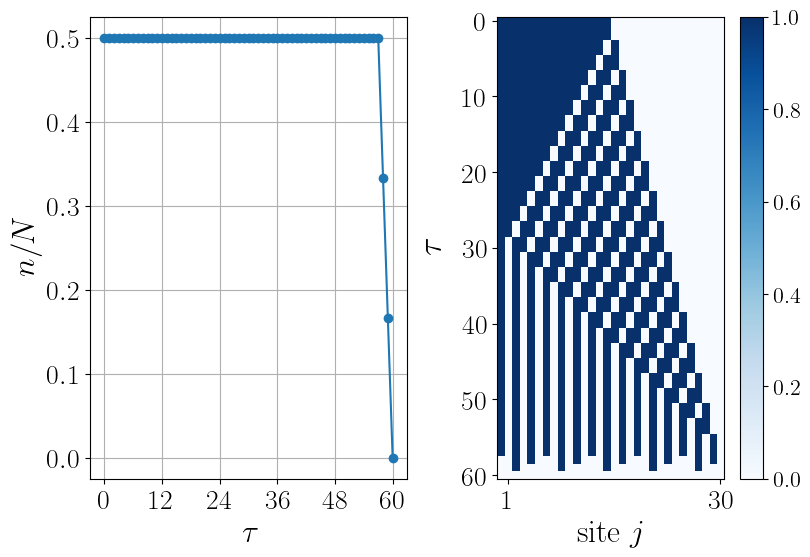}
    \includegraphics[scale=0.4]{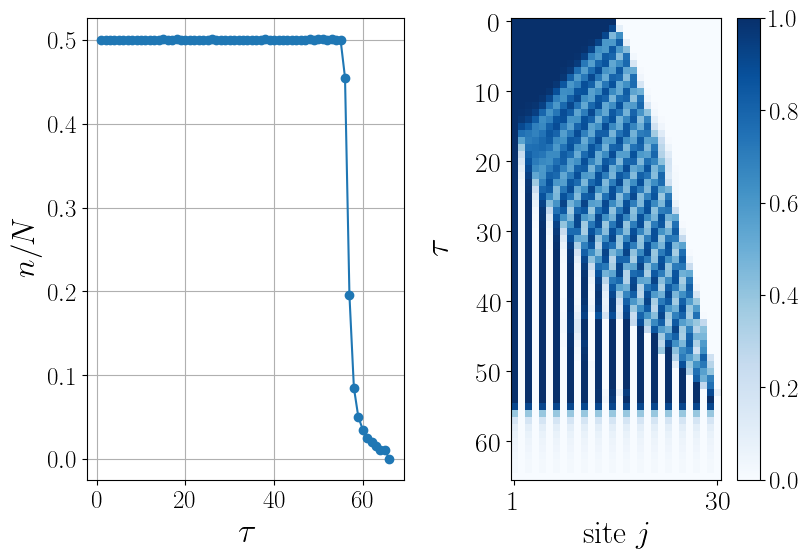}
    \includegraphics[scale=0.4]{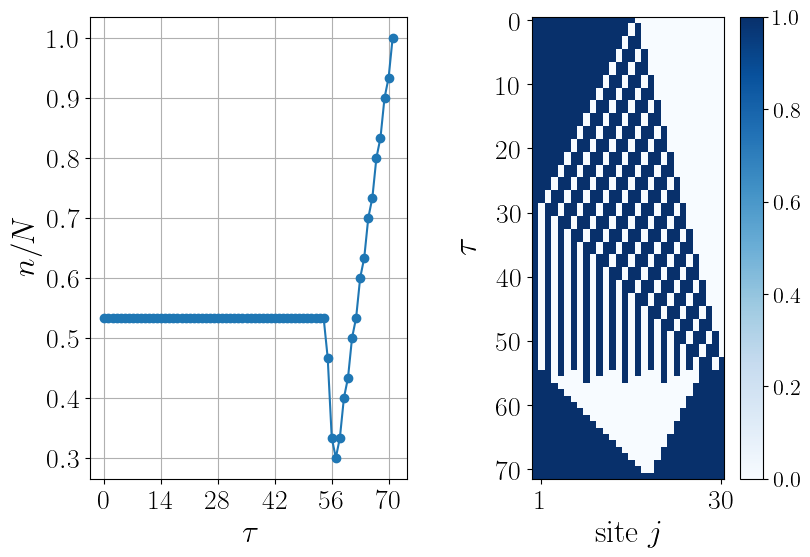}
    \includegraphics[scale=0.4]{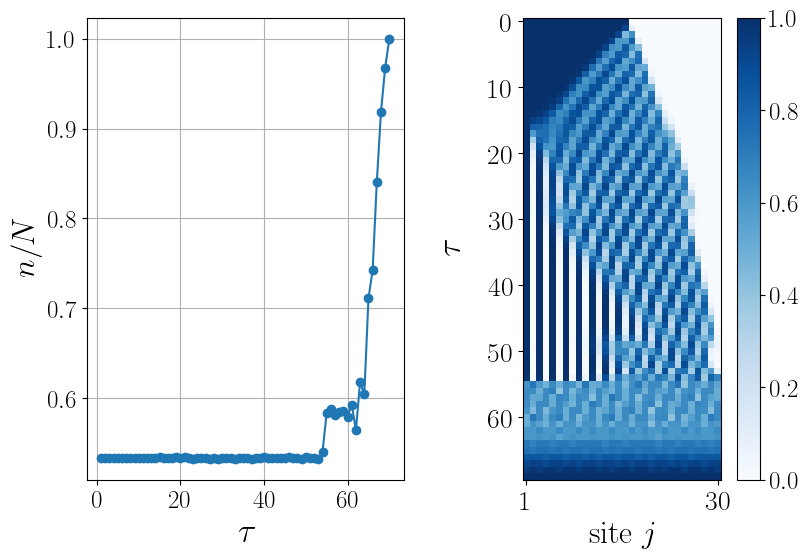}
    \caption{Two examples of how our proposed solution successfully solves the majority voting problem both with discrete-time (left plots) and continuous-time (right plots) evolution, starting from initial states with \( N=30 \) and consisting of 15 \( \ket{1} \)s (top panel) and 16 \( \ket{1} \)s (bottom panel), respectively. In each plot, the variation in \( n/N \) (with $n$ the expectation value of $\hat{n}=\sum_j \hat{P}_{1j}$) as a function of \( \tau \) and the QCA evolution are shown.}
    \label{fig:plots_MV}
\end{figure}

\begin{figure}[h]
    \includegraphics[scale=0.4]{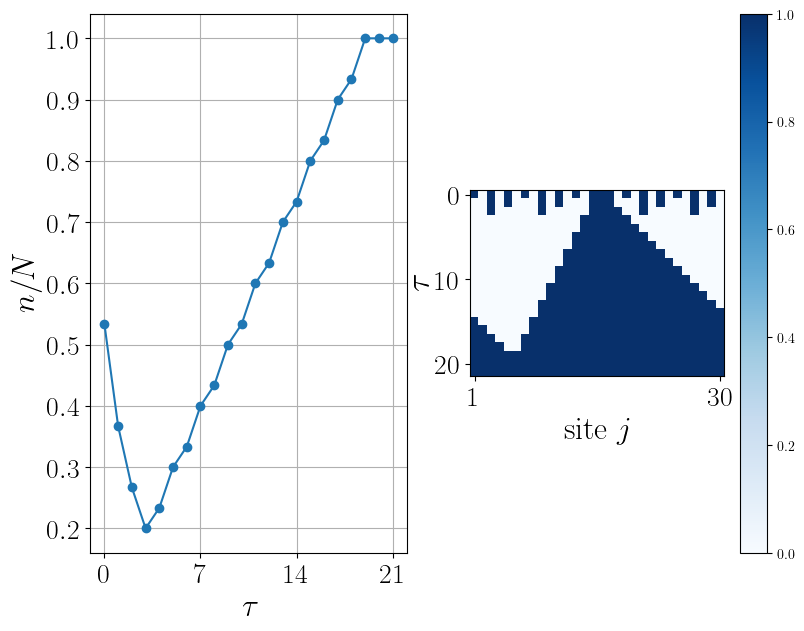}
    \includegraphics[scale=0.4]{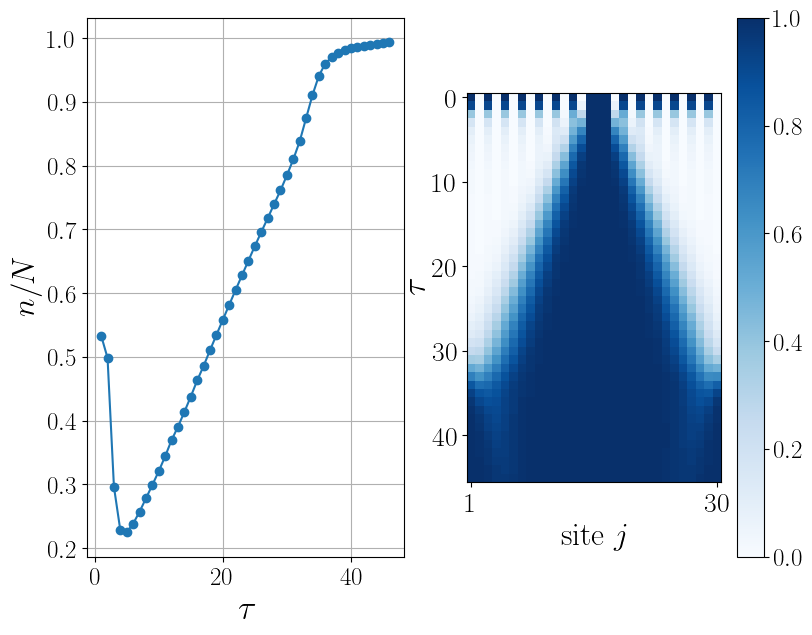}
    \caption{Comparison between discrete-time evolution $\hat{\mathbb{B}}^{\tau}$ (left plots) and continuous-time evolution $e^{\hat{\mathbb{L}}^{\text{B}} \tau}$ (right plots) acting on the input state $\ket{101010101010101110101010101010}$ without applying the corresponding  $\hat{\mathbb{A}}$-type evolution first. Notice how the density momentarily decreases as isolated $1$'s are mapped to $0$'s, while the isolated initial cluster of $1$~s grows and eventually dominates to map the chain to a uniform $1$ state.
    }
    \label{fig:plot_discrete3}
\end{figure}

As outlined in Section~\ref{sec:mv}, a continuous-time evolution proposal is possible (see Equation~\eqref{eq:continuous_time}).
To showcase different scenarios, two initial states, belonging to two different sectors of $n$ (with $n$ the expectation value of $\hat{n}=\sum_j \hat{P}_{1j}$), are chosen in Figure~\ref{fig:plots_MV}.
These states, having a size $N=30$ and containing 15 and 16 $|1\rangle$s, respectively, are evolved by using both discrete-time and continuous-time evolutions. These numerical simulations of the continuous-time evolutions have been obtained by exploiting the Time-Dependent Variational Principle (TDVP) \cite{haegeman2011time, haegeman2016unifying}, implemented in the ITensor library \cite{itensor-r0.3} in 
 \texttt{C\texttt{++}}.
It is noticeable in Figure~\ref{fig:plots_MV} how the action of $\hat{\mathbb{A}}$ separates and disperses the $\ket{1}$s along the chain, resulting in a state where there are no neighboring $\ket{1}$s (the same can be appreciated in the continuous case under the action of \(\mathbb{L}^\text{A}\)).
When $n/N \leq 1/2$, $\hat{\mathbb{A}}$ successfully achieves its goal, and the subsequent action of $\hat{\mathbb{B}}$ enables the attainment of the state \(|0\rangle^{\otimes N}\). However, when \(n/N > 1/2\), at least one small cluster of $\ket{1}$s survives, providing $\hat{\mathbb{B}}$ with the opportunity to propagate it along the entire chain.
This dual action of $\hat{\mathbb{B}}$ is evident in Figure~\ref{fig:plots_MV} (bottom panel, left plot), where it is thus responsible for the momentary decrease in $n/N$.
Such evidence is no longer clearly observable in the corresponding continuous case because, after evolution generated by  $\hat{\mathbb{L}}^\text{A}$, evolution generated by $\hat{\mathbb{L}}^\text{B}$ will act on a mixture of basis states. If we started from a specific basis state, then we could observe its dual action even in the continuous case (as 
 shown in a simple example in Figure~\ref{fig:plot_discrete3} and one can appreciate how the two evolutions are truly similar). 
Indeed, this dual action consists, on one hand, of transforming every $\ket{1}$ in the chain, preceded and followed by a $\ket{0}$, into a $\ket{0}$, and on the other hand, of enlarging every cluster of $\ket{1}$ states.  

\begin{figure}[h]
    \includegraphics[scale=0.5]{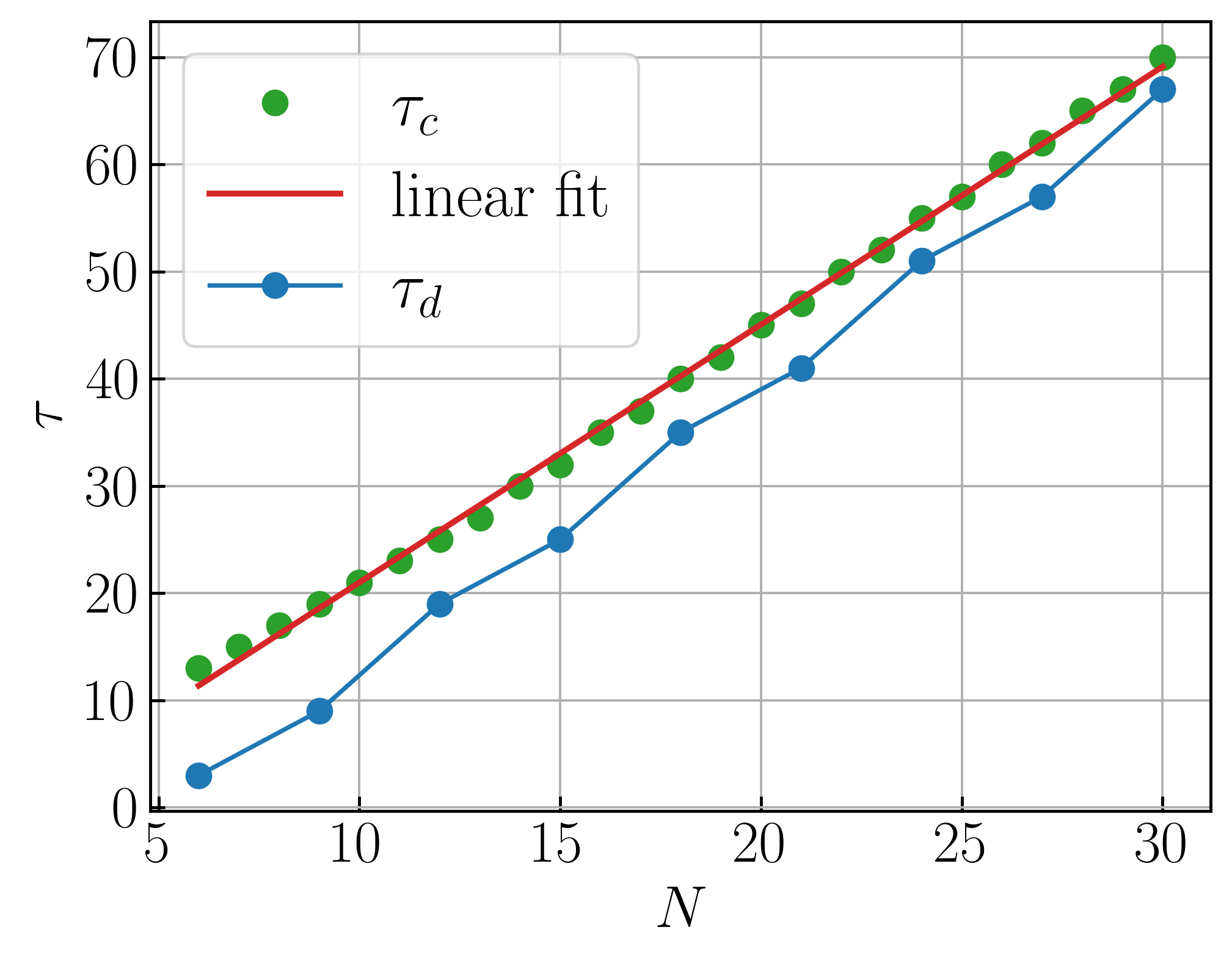}
    \caption{Comparison between the time required to reach the final state of Majority Voting (MV) with discrete-time ($\tau_d$) and continuous-time ($\tau_c$) evolutions, as a function of the system size $N \in [6,30]$. The $\tau_c$ data were computed using the Time-Dependent Variational Principle (TDVP) \cite{haegeman2011time, haegeman2016unifying} implemented in the ITensor library \cite{itensor-r0.3} in \texttt{C\texttt{++}}, whereas the $\tau_d$ data represent the plot of the function~\eqref{eq:scaling_discrete} by selecting $N$mod$(3)=0$. The linear regression fit corresponds to the $\tau_c$ data, yielding $\tau_c(N)=b\times N+q$ with parameters $b=2.40\pm 0.02$ and $q=-3.0 \pm 0.4$.}
    \label{fig:plot_comparison}
\end{figure}

Lastly, we present a comparison of how \(\tau=\tau_A+\tau_B\) scales with the system size $N$ in the discrete- and continuous-time cases; see Figure~\ref{fig:plot_comparison}. In the former, we simply plot Equation~\eqref{eq:scaling_discrete}. Similarly, in the continuous-time case, we compute \(\tau_A\) and \(\tau_B\) in the worst-case scenario: for \(\tau_A\), we consider the desired state achieved when $n/N$ (whose sum, in this case, is taken only over the occupied sites) exceeds 0.99/2, and  for \(\tau_B\), when \(n/N\) exceeds~0.99.

\section{Conclusions} \label{sec:conclusion}
The DC task has been studied using one-dimensional non-unitary QCAs, which perform a computation that maps global information to local information. Two approaches are considered: one that preserves the number density and one that performs majority voting. For the DC, two QCAs have been introduced that have been shown to solve the task by reaching the fixed point with an approximately quadratical time scaling with the system size. One of them is inspired by the \Fuks\ CA \cite{fuks} and the other one is a new quantum model which is restricted to only two-body interactions and has been shown to solve the DC task by a factor of $1.378$ faster than the \Fuks\ QCA. A third QCA model was introduced, which has been shown to solve the majority voting problem within a time that scales linearly with the system size $N$. Both discrete-time CPTP maps as well as the corresponding continuous-time Lindblad dynamics have been considered.

A potential application of our majority voting QCA is for MFQEC mentioned in the introduction for more general noise channels. MFQEC is an alternative to measurement-based QEC suited to architectures where measurements are particularly noisy and slow. The basic approach to MFQEC with stabilizer codes is to map stabilizer outcomes to freshly prepared ancillae using transversal gates, and then to coherently apply correction operations on the data register based on the information contained in the ancillae. For Shor-type MFQEC \cite{MFQEC04}, in order to make it fault tolerant, several repetitions are made of the mapping of stabilizer outcomes to $d$ ancillae, where $d$ is the code distance, and then a majority vote is made on the ancillary register followed by a coherent correction on the data register. Since the ancilla register is in fact quantum, classically processing by a majority voting circuit using Boolean logic is not possible without first translating it into classical data via measurement, which was to be avoided in the first place. The Majority Voting QCA would obviate this by efficiently computing the majority voting in place on the ancilla. Note that other approaches to MFQEC have been proposed including unitary Steane type stabilizer mappings \cite{PRXQuantum.5.010333} and unitary majority voting gadgets \cite{PhysRevLett.105.100501}, but using this QCA approach could simplify some implementations, as it does not require the addressability of the ancillary register.

\section{Acknowledgments}

We acknowledge helpful discussions with Pedro C.~S.~Costa and Yuval R.~Sanders. This research was supported by the Australian Research Council Centre of Excellence for Engineered Quantum Systems (EQUS, CE170100009), and by the Sydney Quantum Academy, Sydney, NSW, Australia, which is supported by the NSW Government (Allocation No.~20201969). Additionally, we acknowledge the support through the project ``QUANTUM'' by the Istituto Nazionale di Fisica Nucleare, Sezione di Bologna, Italy.
E. Wagner and F. Dell'Anna contributed equally to this work.

\bibliography{biblio}

\appendix

\begin{appendix}
\section{Derivation of the Lindbladian Describing the \Fuks\ QCA} \label{app:decayrate}
In this section, it is shown that the relative weightings of the jump operators describing the \Fuks\ QCA are correct (see Equations~\eqref{eq:Fuksjumpop1}--\eqref{eq:Fuksjumpop6}) by establishing a relationship between the probabilities $p$ and the product of the decay rate $\gamma$ and the time duration $\tau$ of each QCA update.
The derivation is based on the idea that the (continuous-time) Lindblad dynamics mimic the (discrete-time) superoperator in Equation~\eqref{eq:s2} with the Kraus operators in Equations~\eqref{eq:krausfuks00}--\eqref{eq:krausfuks4}, i.e.,~the Lindbladian is determined in such a way that it results in the same dynamics as the superoperator for a given time step $\tau$.
Hereby, $\gamma$ is found to have an explicit relation to the probability $p$ of the probabilistic QCA; or in other words, $\gamma$ is found to be tuned in such a way that it implements the QCA for different values of $p\in(0,1/2]$.

The relationship between $\gamma$ and $p$ can be derived by setting the state undergoing the time evolution according to the Lindblad dynamics in Equation~\eqref{eq:LioFuks} equal to the state that is being updated by the discrete-time transfer matrix in Equation~\eqref{eq:s}:
\ba
    e^{\hat{\mathcal{L}}^\tx{(\Fuks)} \tau\,}[\hat\rho]
    =\hat{S}^\tx{(\Fuks)}[\hat\rho].
    \label{eq:equaloperations}
\ea

\subsection{\texorpdfstring{$\ket{00}$}{} Neighborhood}\label{sec:00neighborhood}
For the scope of this proof, it is sufficient to take the evolution of only one qubit into account by fixing the nearest-neighboring qubits to, say at first, the $\dyad{0}_{j-1}\otimes\dyad{0}_{j+1}$ state. In such a way, only the first jump operator $\hat{L}_{1_j} = \sqrt{\gamma}\dyad{0}_{j-1} \otimes\, \hat\sigma^-_{j} \otimes \dyad{0}_{j+1}$ (see Equation~\eqref{eq:Fuksjumpop1}) acts non-trivially on the qubit at site $j$, and the neighboring sites can be traced out for simplicity. 
Analogously, only the Kraus operators 
$\hat K_{0}^\tx{(00)}= \mqty(1&0\\0&\sqrt{1-2p})$, and $\hat K_{1}^\tx{(00)}= \mqty(0&\sqrt{2p}\\0&0)$ (see Equation~\eqref{eq:krausfuks00}) must be taken into account because $\mathbb{\hat{S}}^\tx{(00)}$ in Equation~\eqref{eq:s2} is the only transfer operator that acts on the state at the center site $j$ given the $\dyad{0}_{j-1}\otimes\dyad{0}_{j+1}$ neighborhood.

The corresponding continuous-time evolution \eqref{eq:LioFuks} of the quantum state\linebreak  $\hat\rho=\mqty(\rho_{00}&\rho_{01}\\\rho_{10}&\rho_{11})$ at site $j$ is then given by
\ba
    \hat{\mathcal{L}}^{(00)}[\hat{\rho}]
    &=
    \gamma\qty[\hat{\sigma}^-\hat{\rho}\,\hat\sigma^+ - \frac{1}{2}(\hat\sigma^+\hat\sigma^-\hat{\rho} + \hat\rho\,\hat\sigma^+\hat\sigma^-)] \nn\\[1.5\jot]
    &=
    \gamma \qty[\mqty(0&1\\0&0)\mqty(\rho_{00}&\rho_{01}\\\rho_{10}&\rho_{11})\mqty(0&0\\1&0)
    - 
    \frac{1}{2}\qty(
    \mqty(0&0\\1&0)\mqty(0&1\\0&0)
    \mqty(\rho_{00}&\rho_{01}\\\rho_{10}&\rho_{11})
    +
    \mqty(\rho_{00}&\rho_{01}\\\rho_{10}&\rho_{11})
    \mqty(0&0\\1&0)\mqty(0&1\\0&0)
    )] \nn\\[1.5\jot]
    &=
    -\gamma \mqty(-(1-\rho_{00})&\rho_{01}/2\\[.25em]
    \rho_{10}/2&\rho_{11}),
\ea
which leads with $\rho_{00}=1-\rho_{11}$ to the output state
\ba
    e^{\hat{\mathcal{L}}^{(00)}\tau}[\hat{\rho}]
    = \mqty(1-e^{-\gamma \tau}(1-\rho_{00})&e^{-\gamma \tau/2}\rho_{01}\\[.4em]
    e^{-\gamma \tau/2}\rho_{10}&e^{-\gamma \tau}\rho_{11}).
    \label{eq:rho00l}
\ea

On the other side, the discrete-time evolution \eqref{eq:s2} results in the state
\ba
    \sum_{\mu=0,1} \hat{K}_{\mu}^\tx{(00)} \; \hat\rho \ \qty(\hat{K}_{\mu}^{(00)})^{\!\dagger}
    &=
    \mqty(1&0\\0&\sqrt{1-2p})
    \mqty(\rho_{00}&\rho_{01}\\\rho_{10}&\rho_{11})
    \mqty(1&0\\0&\sqrt{1-2p})
    +
    \mqty(0&\sqrt{2p}\\0&0)
    \mqty(\rho_{00}&\rho_{01}\\\rho_{10}&\rho_{11})
    \mqty(0&0\\\sqrt{2p}&0) \nn\\[2\jot]
    &=
    \mqty((1-2p)\rho_{00}+2p&\sqrt{1-2p}\,\rho_{01}\\[.4em]
    \sqrt{1-2p}\,\rho_{10}&(1-2p)\rho_{11}).
    \label{eq:rho00s}
\ea

Setting the time-evolved density matrices in Equations~\eqref{eq:rho00l} and \eqref{eq:rho00s} equal to each other according to Equation~\eqref{eq:equaloperations}, one can find the relationship between $\gamma$ and $p$ by equating the individual density operator components. For example, taking $\rho_{01}(\tau)$ into account:
\ba
    e^{-\gamma \tau/2} &= \sqrt{1-2p},
\ea
which is equivalent to equating the $\rho_{11}(\tau)$ components and leads to the same result:
\ba
    e^{-\gamma \tau} &= 1-2p \nn\\
    \Rightarrow \gamma \tau &= -\ln\qty(1-2p).
    \label{eq:gammat00}
\ea

Note that the other two density matrix elements exhibit the same information as $\rho_{00}(\tau)=1-\rho_{11}(\tau)$ and $\rho_{10}(\tau)=\qty(\rho_{01}(\tau))^*$.

\subsection{\texorpdfstring{$\ket{01}$}{} Neighborhood}
Next, fixing the nearest-neighboring qubits to the $\dyad{0}_{j-1}\otimes\dyad{1}_{j+1}$ state,
only the two jump operators $\hat{L}_{2_j} = \sqrt{\gamma}\dyad{0}_{j-1} \otimes \,\hat\sigma^-_{j} \otimes \dyad{1}_{j+1}$ and $\hat L_{3_j} = \sqrt{\gamma}\dyad{0}_{j-1} \otimes\, \hat\sigma^+_{j} \otimes \dyad{1}_{j+1}$ in Equations~\eqref{eq:Fuksjumpop2} and \eqref{eq:Fuksjumpop3} as well as the Kraus operators 
$\hat K_{0}^\tx{(01)}=\sqrt{1-p}\ \hat{\mathds{1}}$ and $\hat K_{1}^\tx{(01)}= \sqrt{p}\, \hat{X}$ from Equation~\eqref{eq:krausfuks01} have to be taken into account.
The corresponding Lindblad evolution \eqref{eq:LioFuks} of the quantum state $\hat\rho$ at site $j$ is then given by
\ba
    \hat{\mathcal{L}}^{(01)}[\hat{\rho}]
    &=
    \frac{\gamma}{2}\qty[\hat{\sigma}^-\hat{\rho}\,\hat\sigma^+ +\hat \sigma^+\hat\rho\,\hat\sigma^- - \frac{1}{2}(\hat\sigma^+\hat\sigma^-\hat\rho + \hat\rho\,\hat\sigma^+\hat\sigma^-)
    - \frac{1}{2}(\hat\sigma^-\hat\sigma^+\hat\rho + \hat\rho\,\hat\sigma^-\hat\sigma^+)] \nn\\
    &=
    \frac{\gamma}{2}\Bigg[\mqty(0&1\\0&0)\mqty(\rho_{00}&\rho_{01}\\\rho_{10}&\rho_{11})\mqty(0&0\\1&0)
    +
    \mqty(0&0\\1&0)\mqty(\rho_{00}&\rho_{01}\\\rho_{10}&\rho_{11})\mqty(0&1\\0&0)\nn\\
    &\qquad- \frac{1}{2}
    \qty(
    \mqty(0&0\\1&0)\mqty(0&1\\0&0)
    \mqty(\rho_{00}&\rho_{01}\\\rho_{10}&\rho_{11})
    +
    \mqty(\rho_{00}&\rho_{01}\\\rho_{10}&\rho_{11})
    \mqty(0&0\\1&0)\mqty(0&1\\0&0))\nn\\
    &\qquad- \frac{1}{2}
    \qty(
    \mqty(0&1\\0&0)\mqty(0&0\\1&0)
    \mqty(\rho_{00}&\rho_{01}\\\rho_{10}&\rho_{11})
    +
    \mqty(\rho_{00}&\rho_{01}\\\rho_{10}&\rho_{11})
    \mqty(0&1\\0&0)\mqty(0&0\\1&0))
    \Bigg] \nn\\
&=
    \frac{\gamma}{2} \Bigg[\mqty(\rho_{11}&0\\0&0)
    +\mqty(0&0\\0&\rho_{00})- \frac{1}{2} 
    \qty(
        \mqty(0&0\\\rho_{10}&\rho_{11})
        +\mqty(0&\rho_{01}\\0&\rho_{11})
        +\mqty(\rho_{00}&\rho_{01}\\0&0)
        +\mqty(\rho_{00}&0\\\rho_{10}&0)
    )\Bigg] \nn\\
    &=
    -\gamma \mqty(\rho_{00}-\frac{1}{2}&\rho_{01}/2\\\rho_{10}/2&\rho_{11}-\frac{1}{2}),
\label{eq:L01fuks}
\ea
where the index $j$ is dropped for clarity, and $\rho_{00}=1-\rho_{11}$ is used in the last step. This result leads to the output state
\ba
    e^{\hat{\mathcal{L}}^{(01)}\tau}[\hat{\rho}]
    = \mqty(e^{-\gamma \tau}\rho_{00}+\frac{1}{2}(1-e^{-\gamma \tau})&e^{-\gamma \tau/2}\rho_{01}\\[.5em]
    e^{-\gamma \tau/2}\rho_{10}&e^{-\gamma \tau}\rho_{11}+\frac{1}{2}(1-e^{-\gamma \tau})).
    \label{eq:rho01l}
\ea

On the other side, the discrete-time evolution \eqref{eq:s2} results in the state
\ba
    \sum_{\mu=0,1} \hat{K}_{\mu}^\tx{(01)} \; \hat\rho \ \qty(\hat{K}_{\mu}^{(01)})^{\!\dagger}
    &=
    (1-p)\ \hat{\mathds{1}}
    \mqty(\rho_{00}&\rho_{01}\\\rho_{10}&\rho_{11})
    \hat{\mathds{1}}
    +
    p\mqty(0&1\\1&0)
    \mqty(\rho_{00}&\rho_{01}\\\rho_{10}&\rho_{11})
    \mqty(0&1\\1&0) \nn\\
    &=
    (1-p)    \mqty(\rho_{00}&\rho_{01}\\\rho_{10}&\rho_{11})
    +
    p    \mqty(\rho_{11}&\rho_{10}\\\rho_{01}&\rho_{00})\nn\\
    &=
    \mqty((1-2p)\rho_{00}+p&(1-p)\,\rho_{01}+p\rho_{10}\\[.5em](1-p)\,\rho_{10}+p\rho_{01}&(1-2p)\rho_{11}+p),
    \label{eq:rho01s}
\ea
with (again) $\rho_{00}=1-\rho_{11}$ applied in the last step.
Setting the time-evolved density operators in Equations~\eqref{eq:rho01l} and \eqref{eq:rho01s} equal to each other according to Equation~\eqref{eq:equaloperations}, one can, analogously to the previous subsection in Appendix~\ref{sec:00neighborhood}, find the relationship between $\gamma$ and $p$ by equating the individual density operator components. Taking $\rho_{01}(t)$ into account, one can find that
\ba
    e^{-\gamma \tau/2}\rho_{01} = (1-p)\,\rho_{01}+p\,\rho_{10}
\ea
does not lead to a unique solution; however, equating the $\rho_{11}(t)$ components leads to the same result \eqref{eq:gammat00} derived using the $00$ neighborhood in the previous subsection:
\ba
    e^{-\gamma \tau}\rho_{00}+\frac{1}{2}(1-e^{-\gamma \tau}) &= (1-2p)\rho_{00}+p \nn\\
    \Rightarrow e^{-\gamma \tau} &= 1-2p \nn\\
    \Rightarrow \gamma \tau &= -\ln\qty(1-2p).
    \label{eq:gammatau}
\ea

Equivalent expressions hold for the density matrix components $\rho_{00}(t)=1-\rho_{11}(t)$ and $\qty(\rho_{10}(t)=\rho_{01}(t))^*$, respectively.
Furthermore, the same relationship between $p$ and $\gamma$ follows from an analogous inspection of the $10$ and $11$ neighborhoods.

\section{Proof of the Conservation of the Number Density in the \Fuks\ QCA} \label{app:SzFuks}

In the following, it is derived that the global number density of the \Fuks\ QCA is conserved. The proof arises from the conservation of the expectation value of $\hat{S}_z=\frac{1}{2}\sum_i \hat{Z}_i$, considering the corresponding Lindblad dynamics in Equation~\eqref{eq:LioFuks}. 
The time derivative of $\hat{S}_z(t)$, expressed as the trace of the product of $\hat{S}_z$ and the Lindbladian acting on the density matrix $\hat\rho(t)$, can be shown to be equal to the trace of the product of $\hat\rho(t)$ and the \textit{{adjoint}} Lindbladian $\hat{\mathcal L}^\dagger$ acting on $\hat{S}_z$:
\ba
    \frac{\tx d}{\tx dt}\expval{\hat{S}_z(t)}
    =\Tr \qty[\hat{\mathcal{L}} \qty[\hat\rho(t)] \cdot \hat{S}_z]
    =\Tr \qty[\hat{\mathcal{L}}^\dagger \qty[\hat{S}_z] \cdot \hat\rho(t)].
    \label{eq:dtsz}
\ea

Thus, the problem reduces to the calculation of
\ba
    \hat{\mathcal{L}}^\dagger \qty[\hat{S}_z]
    =\hat{\mathcal{L}}^\dagger \qty[\frac{1}{2}\sum_i \hat{Z}_i]
    =\frac{1}{2} \sum_{i} \hat{\mathcal{L}}^\dagger \qty[\hat{Z}_i]
    =\frac{1}{2} \sum_{i,j} \hat{\mathcal{L}}_j^\dagger \qty[\hat{Z}_i],
    \label{eq:adjointliosum}
\ea
where $\hat{\mathcal{L}}_j^\dagger$ represents the local adjoint Lindbladian that acts non-trivially on the neighborhood at sites $j-1$, $j$, and $j+1$ only; see the definition of the associate jump operators in Equations~\eqref{eq:Fuksjumpop1}--\eqref{eq:Fuksjumpop6}.
This means that 
\ba
    \hat{\mathcal{L}}_j^\dagger \qty[\hat{Z}_i]
    =\sum_{k=1}^6 
    \qty(\hat L_{k_j}^\dagger \hat{Z}_i \hat L_{k_j}- \frac{1}{2} \qty(\hat{Z}_i \hat L_{k_j}^\dagger \hat L_{k_j} + \hat L_{k_j}^\dagger \hat L_{k_j} \hat{Z}_i ))
    \label{eq:adjointlio}
\ea
is only non-zero {if and only if} 
 $i\in\{j-1,j,j+1\}$ because for all other $i$, $\hat{\mathcal{L}}_j^\dagger$ acts merely trivially on $\hat{Z}_i$ \Big(as all jump operators are the identity operator in this case, $\hat L_{k_i}=\hat{\mathds{1}}_i \ \forall\ k\in[1,6]$, such that $\hat{\mathcal{L}}_j^\dagger\qty[\hat{Z}_{i\neq \{j-1,j,j+1\}}]$ only includes terms like $\hat{\mathds{1}}_i^\dagger \hat{Z}_i\hat{\mathds{1}}_i=\hat{Z}_i$\Big); additionally, its action onto the identity operators at sites $\{j-1,j,j+1\}$ returns zero, and $\hat{\mathcal{L}}_j^\dagger \qty[\mathds{\hat{1}}_{j-1}\otimes\hat{\mathds{1}}_j\otimes\hat{\mathds{1}}_{j+1}]=0$, according to Equation~\eqref{eq:adjointlio}.
(Another (potentially more intuitive) way to think about this derivation is to notice that the identity channel possesses a constant expectation value,
\ba
    \frac{\tx d}{\tx dt}\expval{\hat O(t)}
    =\frac{\tx d}{\tx dt} \Tr \qty[\hat O \,\hat\rho(t)]=0
    \tx{ if } \hat O = \mathds{\hat{1}},
\ea
and does therefore not change the total time derivative of the expectation value of $\hat{S}_z$.)

As a result, Equation~\eqref{eq:adjointliosum} simplifies to
\ba
    \hat{\mathcal{L}}^\dagger \qty[\hat{S}_z]
    =\frac{1}{2} \sum_{j} \qty(\hat{\mathcal{L}}_j^\dagger\qty[\hat{Z}_{j-1}]
    +\hat{\mathcal{L}}_j^\dagger\qty[\hat{Z}_{j}]
    +\hat{\mathcal{L}}_j^\dagger\qty[\hat{Z}_{j+1}]),
    \label{eq:adjointlio2}
\ea
where each summand is going to be inspected separately in the following.

To start, $\hat{\mathcal{L}}_j^\dagger\qty[\hat{Z}_{j-1}]$ includes the terms $\sum_{k=1}^6 \hat L_{k_j}^\dagger  \hat{Z}_{j-1} \hat L_{k_j}$, $\sum_{k=1}^6 \hat{Z}_{j-1} \hat L_{k_j}^\dagger \hat L_{k_j}$ and $\sum_{k=1}^6 \hat L_{k_j}^\dagger \hat L_{k_j}\hat{Z}_{j-1}$, see Equation~\eqref{eq:adjointlio}.
The first sum comprises the terms according to the jump operators in Equations~\eqref{eq:Fuksjumpop1}--\eqref{eq:Fuksjumpop6}, where $\gamma\equiv1$ for clarity:\vspace{6pt}
\ba
   \hat{L}_{1_j}^\dagger \hat{Z}_{j-1} \hat{L}_{1_j}
    &=
    \dyad{0} \hat{Z} \dyad{0} \otimes \dyad{1} \otimes \dyad{0}
    &&=
    \dyad{0} \otimes \dyad{1} \otimes \dyad{0}
    \nn\\
   \hat{L}_{2_j}^\dagger \hat{Z}_{j-1} \hat{L}_{2_j}
    &=\frac{1}{2}
    \dyad{0} \hat{Z} \dyad{0} \otimes \dyad{1} \otimes \dyad{1}
    &&=\frac{1}{2}
    \dyad{0} \otimes \dyad{1} \otimes \dyad{1}
    \nn\\
   \hat L_{3_j}^\dagger \hat{Z}_{j-1} \hat L_{3_j}
    &=\frac{1}{2}
    \dyad{0} \hat{Z} \dyad{0} \otimes \dyad{0} \otimes \dyad{1}
    &&=\frac{1}{2}
    \dyad{0} \otimes \dyad{0} \otimes \dyad{1}
    \nn\\
   \hat L_{4_j}^\dagger \hat{Z}_{j-1} \hat L_{4_j}
    &=\frac{1}{2}
    \dyad{1} \hat{Z} \dyad{1} \otimes \dyad{1} \otimes \dyad{0}
    &&=-\frac{1}{2}
    \dyad{1} \otimes \dyad{1} \otimes \dyad{0}
    \nn\\
   \hat L_{5_j}^\dagger \hat{Z}_{j-1} \hat L_{5_j}
    &=\frac{1}{2}
    \dyad{1} \hat{Z} \dyad{1} \otimes \dyad{0} \otimes \dyad{0}
    &&=-\frac{1}{2}
    \dyad{1} \otimes \dyad{0} \otimes \dyad{0}
    \nn\\
   \hat L_{6_j}^\dagger \hat{Z}_{j-1} \hat L_{6_j}
    &=
    \dyad{1} \hat{Z} \dyad{1} \otimes \dyad{0} \otimes \dyad{1}
    &&=
    -\dyad{1} \otimes \dyad{0} \otimes \dyad{1},
\ea
which are the same as the terms in the second sum,
\ba
    \hat{Z}_{j-1} \hat{L}_{1_j}^\dagger \hat{L}_{1_j}
    &=
    \hat{Z} \dyad{0} \dyad{0} \otimes \dyad{1} \otimes \dyad{0}
    &&=
    \dyad{0} \otimes \dyad{1} \otimes \dyad{0}
    \nn\\
    \hat{Z}_{j-1} \hat{L}_{2_j}^\dagger \hat{L}_{2_j}
    &=\frac{1}{2}
    \hat{Z} \dyad{0} \dyad{0} \otimes \dyad{1} \otimes \dyad{1}
    &&=\frac{1}{2}
    \dyad{0} \otimes \dyad{1} \otimes \dyad{1}
    \nn\\
    \hat{Z}_{j-1} \hat L_{3_j}^\dagger \hat L_{3_j}
    &=\frac{1}{2}
    \hat{Z} \dyad{0} \dyad{0} \otimes \dyad{0} \otimes \dyad{1}
    &&=\frac{1}{2}
    \dyad{0} \otimes \dyad{0} \otimes \dyad{1}
    \nn\\
    \hat{Z}_{j-1} \hat L_{4_j}^\dagger \hat L_{4_j}
    &=\frac{1}{2}
    \hat{Z} \dyad{1} \dyad{1} \otimes \dyad{1} \otimes \dyad{0}
    &&=-\frac{1}{2}
    \dyad{1} \otimes \dyad{1} \otimes \dyad{0}
    \nn\\
    \hat{Z}_{j-1} \hat L_{5_j}^\dagger \hat L_{5_j}
    &=\frac{1}{2}
    \hat{Z} \dyad{1} \dyad{1} \otimes \dyad{0} \otimes \dyad{0}
    &&=-\frac{1}{2}
    \dyad{1} \otimes \dyad{0} \otimes \dyad{0}
    \nn\\
    \hat{Z}_{j-1} \hat L_{6_j}^\dagger \hat L_{6_j}
    &=
    \hat{Z} \dyad{1} \dyad{1} \otimes \dyad{0} \otimes \dyad{1}
    &&=
    -\dyad{1} \otimes \dyad{0} \otimes \dyad{1},
\ea

\noindent
and analogously the same as the third due to the symmetric action of the hermitian Pauli operator $\hat{Z}=\hat{Z}^\dagger$ onto the binary basis elements. In total, one can find that $\sum_{k=1}^6 \hat L_{k_j} \hat{Z}_{j-1} \hat L_{k_j}^\dagger=\sum_{k=1}^6 \hat{Z}_{j-1} \hat L_{k_j}^\dagger \hat L_{k_j}=\sum_{k=1}^6 \hat L_{k_j}^\dagger \hat L_{k_j} \hat{Z}_{j-1}$,
such that $\hat{\mathcal{L}}_j^\dagger\qty[\hat{Z}_{j-1}]$ is vanishing according to Equation~\eqref{eq:adjointlio}. In the same way, one can show that $\hat{\mathcal{L}}_j^\dagger\qty[\hat{Z}_{j+1}]=0$ by reflection symmetry around the center site $j$. 

Hence, Equation~\eqref{eq:adjointlio2} becomes
\ba
    \hat{\mathcal{L}}^\dagger \qty[\hat{S}_z]=\frac{1}{2} \sum_{j} \hat{\mathcal{L}}_j^\dagger\qty[\hat{Z}_{j}],
    \label{eq:adjointlio3}
\ea
where $\hat{\mathcal{L}}_j^\dagger\qty[\hat{Z}_{j}]$ includes the summands
\ba
   \hat{L}_{1_j}^\dagger \hat{Z}_{j} \hat{L}_{1_j}
    &=
    \dyad{0} \otimes \dyad{1}{0}Z\dyad{0}{1} \otimes \dyad{0}
    &&=
    \dyad{0} \otimes \dyad{1} \otimes \dyad{0}
    \nn\\
   \hat{L}_{2_j}^\dagger \hat{Z}_{j} \hat{L}_{2_j}
    &=\frac{1}{2}
    \dyad{0} \otimes \dyad{1}{0}Z\dyad{0}{1} \otimes \dyad{1}
    &&=\frac{1}{2}
    \dyad{0} \otimes \dyad{1} \otimes \dyad{1}
    \nn\\
   \hat L_{3_j}^\dagger \hat{Z}_{j} \hat L_{3_j}
    &=\frac{1}{2}
    \dyad{0} \otimes \dyad{0}{1} \hat{Z} \dyad{1}{0} \otimes \dyad{1}
    &&=-\frac{1}{2}
    \dyad{0} \otimes \dyad{0} \otimes \dyad{1}
    \nn\\
   \hat L_{4_j}^\dagger \hat{Z}_{j} \hat L_{4_j}
    &=\frac{1}{2}
    \dyad{1} \otimes \dyad{1}{0}Z\dyad{0}{1} \otimes \dyad{0}
    &&=\frac{1}{2}
    \dyad{1} \otimes \dyad{1} \otimes \dyad{0}
    \nn\\
   \hat L_{5_j}^\dagger \hat{Z}_{j} \hat L_{5_j}
    &=\frac{1}{2}
    \dyad{1} \otimes \dyad{0}{1} \hat{Z} \dyad{1}{0} \otimes \dyad{0}
    &&=-\frac{1}{2}
    \dyad{1} \otimes \dyad{0} \otimes \dyad{0}
    \nn\\
   \hat L_{6_j}^\dagger \hat{Z}_{j} \hat L_{6_j}
    &=
    \dyad{1} \otimes \dyad{0}{1} \hat{Z} \dyad{1}{0} \otimes \dyad{1}
    &&=
    -\dyad{1} \otimes \dyad{0} \otimes \dyad{1}
\ea
and
\ba
    \hat{Z}_{j} \hat{L}_{1_j}^\dagger \hat{L}_{1_j}
    = \hat{L}_{1_j}^\dagger \hat{L}_{1_j} \hat{Z}_{j}
    &=
    \dyad{0} \otimes Z\dyad{1} \otimes \dyad{0}
    &&=
    -\dyad{0} \otimes \dyad{1} \otimes \dyad{0}
    \nn\\
    \hat{Z}_{j} \hat{L}_{2_j}^\dagger \hat{L}_{2_j}
    = \hat{L}_{2_j}^\dagger \hat{L}_{2_j} \hat{Z}_{j}
    &=\frac{1}{2}
    \dyad{0} \otimes Z\dyad{1}\otimes \dyad{1}
    &&=-\frac{1}{2}
    \dyad{0} \otimes \dyad{1} \otimes \dyad{1}
    \nn\\
    \hat{Z}_{j} \hat L_{3_j}^\dagger \hat L_{3_j}
    = \hat L_{3_j}^\dagger \hat L_{3_j} \hat{Z}_{j}
    &=\frac{1}{2}
    \dyad{0} \otimes Z\dyad{0} \otimes \dyad{1}
    &&=\frac{1}{2}
    \dyad{0} \otimes \dyad{0} \otimes \dyad{1}
    \nn\\
    \hat{Z}_{j} \hat L_{4_j}^\dagger \hat L_{4_j}
    = \hat L_{4_j}^\dagger \hat L_{4_j} \hat{Z}_{j}
    &=\frac{1}{2}
    \dyad{1} \otimes Z\dyad{1} \otimes \dyad{0}
    &&=-\frac{1}{2}
    \dyad{1} \otimes \dyad{1} \otimes \dyad{0}
    \nn\\
    \hat{Z}_{j} \hat L_{5_j}^\dagger \hat L_{5_j}
    = \hat L_{5_j}^\dagger \hat L_{5_j} \hat{Z}_{j}
    &=\frac{1}{2}
    \dyad{1} \otimes Z\dyad{0} \otimes \dyad{0}
    &&=\frac{1}{2}
    \dyad{1} \otimes \dyad{0} \otimes \dyad{0}
    \nn\\
    \hat{Z}_{j} \hat L_{6_j}^\dagger \hat L_{6_j}
    = \hat L_{6_j}^\dagger \hat L_{6_j} \hat{Z}_{j}
    &=
    \dyad{1} \otimes Z\dyad{0} \otimes \dyad{1}
    &&=
    \dyad{1} \otimes \dyad{0} \otimes \dyad{1}.
\ea

One can see that $\sum_{k=1}^6 \hat L_{k_j} \hat{Z}_{j} \hat L_{k_j}^\dagger=-\sum_{k=1}^6 \hat{Z}_{j} \hat L_{k_j}^\dagger \hat L_{k_j}$ and $\sum_{k=1}^6 \hat{Z}_{j} \hat L_{k_j}^\dagger \hat L_{k_j}=\sum_{k=1}^6 \hat L_{k_j}^\dagger \hat L_{k_j} \hat{Z}_{j}$ such that
$\hat{\mathcal{L}}_j^\dagger\qty[\hat{Z}_{j}]=2\sum_{k=1}^6 \hat L_{k_j} \hat{Z}_{j} \hat L_{k_j}^\dagger$ according to the definition of the adjoint Lindbladian; see Equation~\eqref{eq:adjointlio}.

Summarizing, this leads with Equations~\eqref{eq:adjointlio} and \eqref{eq:adjointlio3} to 
\vspace{-12pt}

\ba
    \hat{\mathcal{L}}^\dagger \qty[\hat{S}_z]
    &=\frac{1}{2}\sum_j \Big(
    2\dyad{0}_{j-1} \otimes \dyad{1}_j \otimes \dyad{0}_{j+1}
    +\dyad{0}_{j-1} \otimes \dyad{1}_j \otimes \dyad{1}_{j+1}
    +\dyad{1}_{j-1} \otimes \dyad{1}_j \otimes \dyad{0}_{j+1}\nn\\
    &\qquad\qquad\quad -\dyad{0}_{j-1} \otimes \dyad{0}_j \otimes \dyad{1}_{j+1}
    -\dyad{1}_{j-1} \otimes \dyad{0}_j \otimes \dyad{0}_{j+1}
    -2\dyad{1}_{j-1} \otimes \dyad{0}_j \otimes \dyad{1}_{j+1}\Big)\nn\\
    &=\frac{1}{2}\sum_j \Big(
    \dyad{0}_{j-1} \otimes \dyad{1}_j \otimes \dyad{0}_{j+1}
    +\dyad{0}_{j-1} \otimes \dyad{1}_j \otimes \dyad{1}_{j+1}\nn\\
    &\qquad\qquad\quad +\dyad{0}_{j-1} \otimes \dyad{1}_j \otimes \dyad{0}_{j+1}
    +\dyad{1}_{j-1} \otimes \dyad{1}_j \otimes \dyad{0}_{j+1}\nn\\
    &\qquad\qquad\quad -\dyad{0}_{j-1} \otimes \dyad{0}_j \otimes \dyad{1}_{j+1}
    -\dyad{1}_{j-1} \otimes \dyad{0}_j \otimes \dyad{1}_{j+1}\nn\\
    &\qquad\qquad\quad-\dyad{1}_{j-1} \otimes \dyad{0}_j \otimes \dyad{0}_{j+1}
    -\dyad{1}_{j-1} \otimes \dyad{0}_j \otimes \dyad{1}_{j+1}\Big)\nn\\
    &=\frac{1}{2}\sum_j \Big(
    \dyad{0}_{j-1} \otimes \dyad{1}_j \otimes \hat{\mathds{1}}_{j+1}
    +\hat{\mathds{1}}_{j-1} \otimes \dyad{1}_j \otimes \dyad{0}_{j+1}\nn\\
    &\qquad\qquad\quad -\hat{\mathds{1}}_{j-1} \otimes \dyad{0}_j \otimes \dyad{1}_{j+1}
    -\dyad{1}_{j-1} \otimes \dyad{0}_j \otimes \hat{\mathds{1}}_{j+1}\Big)\nn\\
    &=\frac{1}{2}\sum_j \Big(
    \dyad{0}_j \otimes \dyad{1}_{j+1}
    +\dyad{1}_j \otimes \dyad{0}_{j+1}
    -\dyad{0}_j \otimes \dyad{1}_{j+1}
    -\dyad{1}_j \otimes \dyad{0}_{j+1}\Big)\nn\\
    &=0,
\ea

where, in the first step, the first and the last projectors are each split into a sum of two identical summands; writing $2\dyad{0}_{j-1} \otimes \dyad{1}_j \otimes \dyad{0}_{j+1}=\dyad{0}_{j-1} \otimes \dyad{1}_j \otimes \dyad{0}_{j+1}+\dyad{0}_{j-1} \otimes \dyad{1}_j \otimes \dyad{0}_{j+1}$, and analogously for $-2\dyad{1}_{j-1} \otimes \dyad{0}_j \otimes \dyad{1}_{j+1}$.
In the second step, all summands that are written next to each other are combined by identifying two identical projectors acting on the same site, while the sum of orthogonal projectors acting on the third site simplifies to $\dyad{0}+\dyad{1}=\hat{\mathds{1}}$. Because of the space invariance of lattice sites $j$, one can then in the fourth step shift the first and the last terms by one lattice site to the right, $j\rightarrow j+1$, and choose by convention to not  write down the identity channels explicitly. The projectors thus cancel each other out in the last trivial step.

Plugging this result into the right-hand site of the initial Equation~\eqref{eq:dtsz}, one can see~that
\ba
    \frac{\tx d}{\tx dt}\expval{\hat{S}_z(t)} = 0.
\ea

The conservation of $\hat{S}_z$ for the Lindbladian describing the \Fuks\ QCA has thus been~proved.

\section{Steady States of the Lindbladian Describing the \Fuks\ QCA} \label{app:steadystatesfuks}
This section presents a derivation of the steady states of the continuous-time quantum dynamics describing the \Fuks\ QCA. The corresponding Lindbladian is defined by \mbox{Equations}~\eqref{eq:LioFuks} and \eqref{eq:Fuksjumpop1}--\eqref{eq:Fuksjumpop6} in the main text.
The steady states of this system are by definition invariant in time, i.e.,~they satisfy the equation
\ba
    \hat{\mathcal{L}}^\tx{(\Fuks)} [\hat\rho_\tx{ss}]=\frac{\tx d}{\tx dt}\hat\rho_\tx{ss}=0,
    \label{eq:appss}
\ea
whose solutions will be presented in the following Section~\ref{sec:appss1}. Non-steady states that do not satisfy this equation are discussed in Section~\ref{app:appss2}.

\subsection{The Set of Steady States}\label{sec:appss1}
First, the trivial solutions, the pure states $\dyad{0...0}$ and $\dyad{1...1}$, are derived to be steady states by showing that they fulfill Equation~\eqref{eq:appss}, i.e.,
\bs
\ba
    \hat{\mathcal{L}}^\tx{(\Fuks)} [\dyad{0...0}]&=0,\label{eq:appss000a}\\
    \hat{\mathcal{L}}^\tx{(\Fuks)} [\dyad{1...1}]&=0.\label{eq:appss000b}
\ea\label{eq:appss000}\es

\vspace{-12pt}
Inspecting the form of the Lindbladian in Equations~\eqref{eq:LioFuks} and~\eqref{eq:Fuksjumpop1}--\eqref{eq:Fuksjumpop6}, note that it only exhibits one jump operator that acts on the $\dyad{0}_{j-1}\otimes\dyad{0}_{j+1}$ ($\dyad{1}_{j-1}\otimes\dyad{1}_{j+1}$) neighborhood, $\hat L_1$ in Equation~\eqref{eq:Fuksjumpop1} ($\hat L_6$ in Equation~\eqref{eq:Fuksjumpop6}), that includes the amplitude damping (amplitude pumping) channel acting on the center site, $\dyad{0}{1}_j$ ($\dyad{1}{0}_j$). This annihilation (rising) operator destroys the state if, as in the case of the $\dyad{0...0}$ ($\dyad{1...1}$) state, the center qubit is in the same state as its neighboring qubits, writing $\sigma^-\ket{0}=\dyad{0}{1}\ket{0}=0$ ($\sigma^+\ket{1}=\dyad{1}{0}\ket{1}=0$), which is the basic argument on which the following complete derivation of Equations~\eqref{eq:appss000a} and \eqref{eq:appss000b} is based on.

Dividing the Lindbladian into a sum of superoperators acting on a subset of a three-qubit neighborhood $\hat{\mathcal{L}}=\sum_j\hat{\mathcal{L}}_j$, its action on the state $\dyad{0}_{j-1}\otimes\dyad{0}_j\otimes\dyad{0}_{j+1}=:\dyad{000}_j$ yields\vspace{-6pt}

\ba
    \hat{\mathcal{L}}_j^\tx{(\Fuks)} [\dyad{000}_j]
    &=\sum_{k=1}^6\qty(\hat L_{k_j} \dyad{000}_j L_{k_j}^\dagger - \frac{1}{2} \qty(\dyad{000}_j L_{k_j}^\dagger \hat L_{k_j} + \hat L_{k_j}^\dagger \hat L_{k_j} \dyad{000}_j ))\nn\\
    &=\hat{L}_{1_j} \dyad{000}_j \hat{L}_{1_j}^\dagger - \frac{1}{2} \qty(\dyad{000}_j \hat{L}_{1_j}^\dagger \hat{L}_{1_j} + \hat{L}_{1_j}^\dagger \hat{L}_{1_j} \dyad{000}_j )\nn\\
    &=\gamma\,(
    \dyad{0} \dyad{0} \dyad{0} \otimes \dyad{0}{1}\dyad{0}\dyad{1}{0} \otimes \dyad{0} \dyad{0} \dyad{0} \nn\\
    &\qquad\qquad - \frac{1}{2} (\dyad{0}\dyad{0}\dyad{0} \otimes (\dyad{0}\dyad{1}{0}\dyad{0}{1}+\dyad{1}{0}\dyad{0}{1}\dyad{0}) \otimes \dyad{0}\dyad{0}\dyad{0} ))_j\nn\\
    &=0,
    \label{eq:fukslioapp000}
\ea
and thus, $\hat{\mathcal{L}}[\dyad{0...0}]=0$. Analogously, $\hat{\mathcal{L}} [\dyad{1...1}]=0$ can be shown by replacing $\hat L_1$ with $\hat L_6$ in the second step, and swapping the annihilation and creation operators $\dyad{0}{1}$ and $\dyad{1}{0}$ in the third step.
As linear combinations of steady states are also steady states, all mixed states of the $\dyad{0...0}$ and $\dyad{1...1}$ states are steady states of the system as~well. 

Furthermore, it can be shown that the associate coherence terms remain invariant in this system,
\bs
\ba
    \hat{\mathcal{L}}^\tx{(\Fuks)} [\dyad{0...0}{1...1}]&=0,\label{eq:appss000111a}\\
    \hat{\mathcal{L}}^\tx{(\Fuks)} [\dyad{1...1}{0...0}]&=0,\label{eq:appss000111b}
\ea\label{eq:appss000111}\es
because the projectors included in the jump operators that determine the states of the neighboring sites $j-1$ and $j+1$ (see Equations~\eqref{eq:Fuksjumpop1}--\eqref{eq:Fuksjumpop6}) annihilate all off-diagonal density matrix elements. For example, the action of the first and second jump operators onto the state $\dyad{0}{1}_{j-1}\otimes\dyad{0}{1}_j\otimes\dyad{0}{1}_{j+1}=:\dyad{000}{111}_j$ lead to
\vspace{-6pt}

\bs
\ba
    &\qty(\hat{L}_{1_j}\dyad{000}{111}_j \hat{L}_{1_j}^\dagger - \frac{1}{2} \qty(\dyad{000}{111}_j \hat{L}_{1_j}^\dagger \hat{L}_{1_j} + \hat{L}_{1_j}^\dagger \hat{L}_{1_j} \dyad{000}{111}_j ))\nn\\
    &\quad=\gamma\,(
    \dyad{0} \dyad{0}{1} \dyad{0} \otimes \dyad{0}{1}\dyad{0}{1}\dyad{1}{0} \otimes \dyad{0} \dyad{0}{1} \dyad{0} \nn\\
    &\qquad\qquad - \frac{1}{2} (\dyad{0}{1}\dyad{0} \otimes \dyad{0}{1}\dyad{1}{0}\dyad{0}{1} \otimes \dyad{0}{1}\dyad{0} + \dyad{0}\dyad{0}{1} \otimes \dyad{1}{0}\dyad{0}{1}\dyad{0}{1} \otimes \dyad{0}\dyad{0}{1})  ))_j\nn\\
    &\quad=0,
\ea   \vspace{-18pt}
\ba
    &\qty(\hat{L}_{2_j} \dyad{000}{111}_j \hat{L}_{2_j}^\dagger - \frac{1}{2} \qty(\dyad{000}{111}_j \hat{L}_{2_j}^\dagger \hat{L}_{2_j} + \hat{L}_{2_j}^\dagger \hat{L}_{2_j} \dyad{000}{111}_j ))\nn\\
    &=\gamma\,(
    \dyad{0} \dyad{0}{1} \dyad{0} \otimes \dyad{0}{1}\dyad{0}{1}\dyad{1}{0} \otimes \dyad{1} \dyad{0}{1} \dyad{1} \nn\\
    &\qquad\qquad - \frac{1}{2} (\dyad{0}{1}\dyad{0} \otimes \dyad{0}{1}\dyad{1}{0}\dyad{0}{1} \otimes \dyad{0}{1}\dyad{1} + \dyad{0}\dyad{0}{1} \otimes \dyad{1}{0}\dyad{0}{1}\dyad{0}{1} \otimes \dyad{1}\dyad{0}{1})  ))_j\nn\\
    &=0.
\ea
\es

Hence,
\ba
    \hat\rho_\tx{ss}^\tx{(\Fuks)}=\alpha\dyad{0...0}+\beta\dyad{0...0}{1...1}+\beta^*\dyad{1...1}{0...0}+(1-\alpha)\dyad{1...1}
    \label{eq:appssfinal}
\ea
has been shown to be a set of steady states. The set includes the GHZ state $\frac{1}{2}(\dyad{0...0}+\dyad{0...0}{1...1}+\dyad{1...1}{0...0}+\dyad{1...1})$ with $\alpha=\beta=\frac{1}{2}$, as well as the pure states $\dyad{0...0}$ and $\dyad{1...1}$ with $\alpha=1$ or $\alpha=0$, respectively, and the mixed state $\alpha\dyad{0...0}+(1-\alpha)\dyad{1...1})$ with $\beta=0$ and $\alpha\in(0,1)$, according to Equations~\eqref{eq:appss000a}, \eqref{eq:appss000b}, \eqref{eq:appss000111a} and \eqref{eq:appss000111b}.

Note that all steady states are translationally invariant, which means that the state exhibits the same \textit{local} number density at every lattice site which is (thus) equal to the \textit{global} number density of the whole state. This number density is the same as the global density of the input state $\hat\rho$ as shown in Appendix~\ref{app:SzFuks}.

Explicitly, the amplitude $\alpha$ that defines the number density of the steady state \eqref{eq:appssfinal} is determined by the input state $\hat\rho$ as follows:
\bs
\ba
    \alpha &= \Tr\qty[\hat P_0\;\hat\rho\,],\\
    1-\alpha &= \Tr\qty[\hat P_1\; \hat\rho\,],
\ea
\es
where $\hat P_0=\sum_j \frac{\hat{\mathds{1}}_j+\hat Z_j}{2}$ and $\hat P_1=\sum_j \frac{\hat{\mathds{1}}_j-\hat Z_j}{2}$. As a simple example, the map evolves the input state $\hat\rho=\dyad{0001}{0001}$ to the steady state $\frac{3}{4}\dyad{0000}{0000}+\frac{1}{4}\dyad{1111}{1111}$, where $\alpha=\frac{3}{4}$ defines the normalized number of zero states and $1-\alpha=\frac{1}{4}$ the normalized number of one states of the input state. The amplitudes $\beta$ and $\beta^*$ of the off-diagonal coherence terms in~\eqref{eq:appssfinal} are given by
\bs
\ba
    \beta &= \Tr\qty[\dyad{0...0}{1...1}\,\hat\rho\,],\\[1.2\jot]
    \beta^* &= \Tr\qty[\dyad{1...1}{0...0}\,\hat\rho\,],
    \label{eq:betabetastar}
\ea\es
with upper bound $|\beta| \leq \sqrt{\alpha(1-\alpha)}$. 

For the proof of Equation~\eqref{eq:betabetastar}, we are going to investigate whether a state $\hat\rho$ that does not exhibit the density matrix element $\dyad{0...0}{1...1}$ (or its complex conjugate) would evolve into a state that does include the density matrix element $\dyad{0...0}{1...1}$ state (or its complex conjugate) under long-time evolution, i.e.,~if that would be the case, then \vspace{6pt}
\ba
    \Tr\qty[e^{\hat{\mathcal{L}}^\tx{(\Fuks)}t}[\hat\rho]\,\dyad{1...1}{0...0}\,] =0
\ea
would be satisfied, which is going to be checked in the following. To start, it is sufficient to consider short-time evolution by taking into account that if the state decays in the long-term limit, it does so in the short-term limit too. The Lindblad evolution can then for $t\ll1$ be approximated by
\ba
    e^{\hat{\mathcal{L}}^\tx{(\Fuks)}t}\qty[\hat\rho] \approx \hat\rho + t \,\hat{\mathcal{L}}^\tx{(\Fuks)}\qty[\hat\rho],
\ea
where only the last term $\hat{\mathcal{L}}\qty[\hat\rho]$ has the potential to include the $\dyad{0...0}{1...1}$ matrix element, as $\hat\rho$ does not by definition.
Considering the evolution of the density matrix element $\dyad{0...0}{1...1_{j-1}0_{j}1_{j+1}...1}$, where, as the $\dyad{0...0}$ and the $\dyad{1...1}$ states have already been shown to vanish under the action the Lindbladian, see Equations~\eqref{eq:appss000a} and \eqref{eq:appss000b}, only the remaining density matrix elements need to be inspected. Only those terms of the Lindbladian that act on the qubit at site $j$ (i.e.,~the state that is not equal to the states of the surrounding qubits in the lattice) can be non-zero; writing with $\dyad{x}{\tilde x}_{j-1}\otimes\dyad{y}{\tilde y}_j\otimes\dyad{z}{\tilde z}_{j+1}=:\dyad{xyz}{\tilde x\tilde y\tilde z}_j$ $\forall\,x,\tilde x,y,\tilde y,z,\tilde z\in \{0,1\}$:
\ba
    \hat{\mathcal{L}}_{j-1}^\tx{(\Fuks)} [\dyad{000}{110}_{j-1}]
    +\hat{\mathcal{L}}_j^\tx{(\Fuks)} [\dyad{000}{101}_j]
    +\hat{\mathcal{L}}_{j+1}^\tx{(\Fuks)} [\dyad{000}{011}_{j+1}].
\ea

As the local Lindbladians acting on the three-cell neighborhood are
\vspace{-12pt}

\bs
\ba
    \hat{\mathcal{L}}_{j-1} \qty[\dyad{000}{110}_{j-1}]
    &=\sum_{k=1}^6 \qty(\underbrace{\hat L_{k_{j-1}} \dyad{000}{110}_{j-1} \hat L_{k_{j-1}}^\dagger}_{=0\ \forall k} - \frac{1}{2} \Big(\dyad{000}{110}_{j-1} \hat L_{4_{j-1}}^\dagger \hat L_{4_{j-1}}+\underbrace{L_{1_{j-1}}^\dagger \hat L_{1_{j-1}} \dyad{000}{110}_{j-1}}_{=0} \Big))\nn\\
    &=-\frac{\gamma}{4}
    \dyad{000}{110}_{j-1},
\\[2\jot]
    \hat{\mathcal{L}}_j \qty[\dyad{000}{101}_j]
    &=\sum_{k=1}^6 \qty(\underbrace{\hat L_{k_{j}} \dyad{000}{101}_{j} \hat L_{k_{j}}^\dagger}_{=0\ \forall k} - \frac{1}{2} \Big(\dyad{000}{101}_{j} \hat L_{6_{j}}^\dagger \hat L_{6_{j}}+\underbrace{L_{1_{j}}^\dagger \hat L_{1_{j}} \dyad{000}{101}_{j}}_{=0} \Big))\nn\\
    &=-\frac{\gamma}{4}
    \dyad{000}{101}_{j},
\\
    \hat{\mathcal{L}}_{j+1} \qty[\dyad{000}{011}_{j+1}]
    &=\sum_{k=1}^6 \qty(\underbrace{\hat L_{k_{j+1}} \dyad{000}{011}_{j+1} \hat L_{k_{j+1}}^\dagger}_{=0\ \forall k} - \frac{1}{2} \Big(\dyad{000}{011}_{j+1} \hat L_{2_{j+1}}^\dagger \hat L_{2_{j+1}}+\underbrace{L_{1_{j+1}}^\dagger \hat L_{1_{j+1}} \dyad{000}{011}_{j+1}}_{=0} \Big))\nn\\
    &=-\frac{\gamma}{4}
    \dyad{000}{011}_{j+1},
\ea\label{eq:appss010v1d}\es

the Lindbladian does not map the state $\dyad{0...0}{1...1_{j-1}0_{j}1_{j+1}...1}$ to the $\dyad{0...0}{1...1}$ state, where analogous derivations hold for all other off-diagonal elements of the initial state that do not equal the $\dyad{0...0}{1...1}$ state. It has hence been shown that $\beta$ in Equation~\eqref{eq:betabetastar} does indeed represent the amplitude of the density matrix element $\dyad{0...0}{1...1}$ of the initial state because there is no other density matrix element that evolves to this state.

\subsection{States That Are Not Steady States}\label{app:appss2}

How about density matrices that include neighboring sites exhibiting different quantum states---could those be also steady states of the explored system? For gleaning this, a superposition state is considered consisting of an arbitrary convex sum of projectors with all qubits except one (at an arbitrary site $j$) occupying the same state:\vspace{6pt}
\ba
    \hat\rho
    &= a\dyad{0...0}
    + b\dyad{0...0_{j-1}1_j0_{j+1}...0} \nn\\[1.2\jot]
    &\qquad+ c\dyad{1...1_{j-1}0_j1_{j+1}...1}
    + d\dyad{1...1}\nn\\[1.2\jot]
    &\qquad+ e\dyad{0...0}{0...0_{j-1}1_j0_{j+1}...0}
    + e^*\dyad{0...0_{j-1}1_j0_{j+1}...0}{0...0} \nn\\[1.2\jot]
    &\qquad+ f\dyad{1...1}{1...1_{j-1}0_j1_{j+1}...1}
    + f^*\dyad{1...1_{j-1}0_j1_{j+1}...1}{1...1},
    \label{eq:rhossapp010}
\ea
where $d=1-a-b-c$ due to the trace-preserving condition of quantum states.

As the $\dyad{0...0}$ and the $\dyad{1...1}$ states have already been shown to vanish under the action Lindbladian (see Equations~\eqref{eq:appss000a} and \eqref{eq:appss000b}), only the remaining density matrix elements need to be inspected in the following:
\ba
    \hat{\mathcal{L}}^\tx{(\Fuks)}[\hat\rho]
    =b\,&\hat{\mathcal{L}}^\tx{(\Fuks)}[\dyad{0...0_{j-1}1_j0_{j+1}...0}]
    +c\,\hat{\mathcal{L}}^\tx{(\Fuks)}[\dyad{1...1_{j-1}0_j1_{j+1}...1}]\nn\\
    &+e\,\hat{\mathcal{L}}^\tx{(\Fuks)}[\dyad{0...0}{0...0_{j-1}1_j0_{j+1}...0}]
    +e^*\,\hat{\mathcal{L}}^\tx{(\Fuks)}[\dyad{0...0_{j-1}1_j0_{j+1}...0}{0...0}]\nn\\
    &+f\,\hat{\mathcal{L}}^\tx{(\Fuks)}[\dyad{1...1}{1...1_{j-1}0_j1_{j+1}...1}]
    +f^*\,\hat{\mathcal{L}}^\tx{(\Fuks)}[\dyad{1...1_{j-1}0_j1_{j+1}...1}{1...1}],
\label{eq:appss010}
\ea

where, by the similar argument that $\hat{\mathcal{L}}_j[\dyad{000}_j]=0$ (see Equation~\eqref{eq:fukslioapp000}), only those terms of the Lindbladian that act on the qubit at site $j$ (i.e.,~the state that is not equal to the states of the surrounding qubits in the lattice) are non-zero---for example, for the first term in Equation~\eqref{eq:appss010}, this means
\ba
    \hat{\mathcal{L}}_{j-1} [\dyad{001}_{j-1}]
    +\hat{\mathcal{L}}_j [\dyad{010}_j]
    +\hat{\mathcal{L}}_{j+1} [\dyad{100}_{j+1}].
    \label{eq:appss010v1}
\ea

For the derivation of these three terms, it is convenient to notice that only one of the six jump operators in Equations~\eqref{eq:Fuksjumpop1}--\eqref{eq:Fuksjumpop6} act on a given state. In the second term, $\dyad{010}_j$, only the jump operator $\hat{L}_{1_j}$ acts on the state, as the neighborhood is in the state $\dyad{0}_{j-1}\otimes\dyad{0}_{j+1}$ (see Equation~\eqref{eq:Fuksjumpop1}), whereas jump operators $\hat L_3$ and $\hat L_5$ each act on the states of the first and the third terms, $\dyad{001}_{j-1}$ and $\dyad{100}_{j+1}$, due to the respective $\dyad{0}_{j-2}\otimes\dyad{1}_j$ and $\dyad{1}_j\otimes\dyad{0}_{j+2}$ neighborhoods, and the center site being in the $\dyad{0}$ state.
In such a way, the summands in Equation~\eqref{eq:appss010v1} yield
\bs
\ba
    \hat{\mathcal{L}}_{j-1}^\tx{(\Fuks)} \qty[\dyad{001}_{j-1}]
    &=\hat L_{3_{j-1}} \dyad{001}_{j-1} \hat L_{3_{j-1}}^\dagger - \frac{1}{2} \qty(\dyad{001}_{j-1} \hat L_{3_{j-1}}^\dagger \hat L_{3_{j-1}} + \hat L_{3_{j-1}}^\dagger \hat L_{3_{j-1}} \dyad{001}_{j-1} )\nn\\
    &=\frac{\gamma}{2}
    \dyad{0}_{j-2} \otimes \qty(\dyad{1}{0}\dyad{0}\dyad{0}{1}
    -\frac{1}{2}(\dyad{0}\dyad{0}{1}\dyad{1}{0}+\dyad{0}{1}\dyad{1}{0}\dyad{0}))_{\!\!j-1} \otimes \dyad{1}_j \nn\\
    &=-\frac{\gamma}{2} \dyad{0}_{j-2} \otimes (\dyad{0}-\dyad{1})_{j-1} \otimes \dyad{1}_j, \label{eq:appss010v1dx1}
\\[2\jot]
    \hat{\mathcal{L}}_j^\tx{(\Fuks)} \qty[\dyad{010}_j]
    &=\hat{L}_{1_j} \dyad{010}_j \hat{L}_{1_j}^\dagger - \frac{1}{2} \qty(\dyad{010}_j \hat{L}_{1_j}^\dagger \hat{L}_{1_j} + \hat{L}_{1_j}^\dagger \hat{L}_{1_j} \dyad{010}_j )\nn\\
    &=\gamma
    \dyad{0}_{j-1} \otimes \qty(\dyad{0}{1}\dyad{1}\dyad{1}{0}-\frac{1}{2} (\dyad{1}\dyad{1}{0}\dyad{0}{1}+\dyad{1}{0}\dyad{0}{1}\dyad{1}))_{\!\!j} \otimes \dyad{0}_{j+1} \nn\\
    &=\gamma \dyad{0}_{j-1} \otimes (\dyad{0}-\dyad{1})_j \otimes \dyad{0}_{j+1},\label{eq:appss010v1dx2}
\\[2\jot]
    \hat{\mathcal{L}}_{j+1}^\tx{(\Fuks)} \qty[\dyad{100}_{j+1}]
    &=\hat L_{5_{j+1}} \dyad{100}_{j+1} \hat L_{5_{j+1}}^\dagger - \frac{1}{2} \qty(\dyad{100}_{j+1} \hat L_{5_{j+1}}^\dagger \hat L_{5_{j+1}} + \hat L_{5_{j+1}}^\dagger \hat L_{5_{j+1}} \dyad{100}_{j+1} )\nn\\
    &=\frac{\gamma}{2}
    \dyad{1}_j \otimes \qty(\dyad{1}{0}\dyad{0}\dyad{0}{1}
    -\frac{1}{2}(\dyad{0}\dyad{0}{1}\dyad{1}{0}+\dyad{0}{1}\dyad{1}{0}\dyad{0}))_{\!\!j+1} \otimes \dyad{0}_{j+2} \nn\\
    &=-\frac{\gamma}{2} \dyad{1}_j \otimes (\dyad{0}-\dyad{1})_{j+1} \otimes \dyad{0}_{j+2}, \label{eq:appss010v1dx3}
\ea\label{eq:appss010v1dx}\es
so that the overall sum of $\hat{\mathcal{L}}_{j-1} \qty[\dyad{001}_{j-1}]$, $\hat{\mathcal{L}}_j \qty[\dyad{010}_j]$ and $\hat{\mathcal{L}}_{j+1} \qty[\dyad{100}_{j+1}]$ is~non-zero,

\ba
    &\hat{\mathcal{L}}^\tx{(\Fuks)}[\dyad{0...0_{j-1}1_j0_{j+1}...0}] \nn\\
    &=-\frac{\gamma}{2} \dyad{0}_1 \otimes...\otimes \dyad{0}_{j-2} \otimes 
    \Big(
    (\dyad{0}-\dyad{1})_{j-1} \otimes \dyad{1}_j \otimes \dyad{0}_{j+1}\nn\\
    &\qquad\qquad\qquad\qquad\qquad\qquad\qquad\quad
    -
    2\dyad{0}_{j-1} \otimes (\dyad{0}-\dyad{1})_j \otimes \dyad{0}_{j+1} \nn\\
    &\qquad\qquad\qquad\qquad\qquad\qquad\qquad\quad+
    \dyad{0}_{j-1} \otimes \dyad{1}_j \otimes (\dyad{0}-\dyad{1})_{j+1} 
    \Big) \otimes \dyad{0}_{j+2} \otimes...\otimes \dyad{0}_N
\nn\\
    &=-\frac{\gamma}{2} \dyad{0}_1 \otimes...\otimes \dyad{0}_{j-2} \otimes 
    \Big(
    -2\dyad{000}_j
    +4\dyad{010}_j\nn\\
    &\qquad\qquad\qquad\qquad\qquad\qquad\qquad\quad
    -\dyad{011}_j
    -\dyad{110}_j
    \Big) \otimes \dyad{0}_{j+2} \otimes...\otimes \dyad{0}_N
\nn\\
    &=\gamma \dyad{0}_1 \otimes...\otimes \dyad{0}_{j-2} \otimes 
    \Big(
    \dyad{000}_j
    -2\dyad{010}_j\nn\\
    &\qquad\qquad\qquad\qquad\qquad\qquad\qquad\quad
    +\frac{1}{2}(\dyad{011}_j
    +\dyad{110}_j)
    \Big) \otimes \dyad{0}_{j+2} \otimes...\otimes \dyad{0}_N
    \neq0.
    \label{eq:appss010final}
\ea

This result also implies that the associated bit-flipped state  ({$\hat{X}^{(N)}\dyad{0...0_{j-1}1_j0_{j+1}...0}\hat{X}^{(N)}$} $=\dyad{1...1_{j-1}0_j1_{j+1}...1}$ with $\hat{X}^{(N)}:=\hat{X}_1\otimes...\otimes \hat{X}_N$, where $N$ is the number of qubits) is also not a steady state,
\ba
    \hat{\mathcal{L}}^\tx{(\Fuks)}[\dyad{1...1_{j-1}0_j1_{j+1}...1}] 
    =\hat{X}^{(N)}\hat{\mathcal{L}}^\tx{(\Fuks)}[\dyad{0...0_{j-1}1_j0_{j+1}...0}]\hat{X}^{(N)}
    \neq 0,
\ea

because of the symmetric definition of the jump operators with $\hat{X}^{(3)}\hat{L}_1\hat{X}^{(3)}=\hat{L}_6$,{$\hat{X}^{(3)}\hat{L}_2\hat{X}^{(3)}=\hat{L}_3$, and $\hat{X}^{(3)}\hat{L}_4\hat{X}^{(3)}=\hat{L}_5$, where $\hat{X}^{(3)}:=(\hat{X}\otimes \hat{X}\otimes \hat{X})$; see \mbox{Equations~\eqref{eq:Fuksjumpop1}--\eqref{eq:Fuksjumpop6}.}}

Next, the off-diagonal density matrix elements in Equation~\eqref{eq:appss010} are taken into account.

For the first term, $\hat{\mathcal{L}}^\tx{(\Fuks)}[\dyad{0...0}{0...0_{j-1}1_j0_{j+1}...0}]$, the non-zero terms are, analogous to the derivation in Equations~\eqref{eq:appss010v1dx1}--\eqref{eq:appss010v1dx3}:
\vspace{-12pt}

\bs
\ba
    \hat{\mathcal{L}}_{j-1}^\tx{(\Fuks)} \qty[\dyad{000}{001}_{j-1}]
    &=\hat L_{3_{j-1}} \dyad{000}{001}_{j-1} \hat L_{3_{j-1}}^\dagger - \frac{1}{2} \qty(\dyad{000}{001}_{j-1} \hat L_{3_{j-1}}^\dagger \hat L_{3_{j-1}} + \hat L_{3_{j-1}}^\dagger \hat L_{3_{j-1}} \dyad{000}{001}_{j-1} ))
    \nn\\
    &=\frac{\gamma}{2} \dyad{0}_{j-2} \otimes \Big(
    \dyad{1}{0}\dyad{0}\dyad{0}{1}_{j-1} \otimes \dyad{1}\dyad{0}{1}\dyad{1}_{j} 
    \nn\\
    &\qquad-\frac{1}{2}(\dyad{0}\dyad{0}{1}\dyad{1}{0}_{j-1} \otimes \dyad{0}{1}\dyad{1}_{j} + \dyad{0}{1}\dyad{1}{0}\dyad{0}_{j-1} \otimes \dyad{1}\dyad{0}{1}_{j})  \Big)\nn\\
    &=-\frac{\gamma}{4} \dyad{0}_{j-2} \otimes \dyad{0}_{j-1} \otimes \dyad{0}{1}_{j},
\\[2\jot]
    \hat{\mathcal{L}}_j^\tx{(\Fuks)} \qty[\dyad{000}{010}_j]
    &=\hat{L}_{1_j} \dyad{000}{010}_j \hat{L}_{1_j}^\dagger - \frac{1}{2} \qty(\dyad{000}{010}_j \hat{L}_{1_j}^\dagger \hat{L}_{1_j} + \hat{L}_{1_j}^\dagger \hat{L}_{1_j} \dyad{000}{010}_j )\nn\\
    &=\gamma
    \dyad{0}_{j-1} \otimes \qty(\dyad{0}{1}\dyad{0}{1}\dyad{1}{0}-\frac{1}{2} (\dyad{0}{1}\dyad{1}{0}\dyad{0}{1}+\dyad{1}{0}\dyad{0}{1}\dyad{0}{1}))_{\!\!j} \otimes \dyad{0}_{j+1} \nn\\
    &=-\frac{\gamma}{2} \dyad{0}_{j-1} \otimes \dyad{0}{1}_j \otimes \dyad{0}_{j+1},
\\[2\jot]
    \hat{\mathcal{L}}_{j+1}^\tx{(\Fuks)} \qty[\dyad{000}{100}_{j+1}]
    &=\hat L_{5_{j+1}} \dyad{000}{100}_{j+1} \hat L_{5_{j+1}}^\dagger - \frac{1}{2} \qty(\dyad{000}{100}_{j+1} \hat L_{5_{j+1}}^\dagger \hat L_{5_{j+1}} + \hat L_{5_{j+1}}^\dagger \hat L_{5_{j+1}} \dyad{000}{100}_{j+1} )\nn\\
    &=\frac{\gamma}{2}\Big(
    \dyad{1}\dyad{0}{1}\dyad{1}_{j} \otimes \dyad{1}{0}\dyad{0}\dyad{0}{1}_{j+1} \nn\\
    &\qquad-\frac{1}{2}(\dyad{0}{1}\dyad{1}_{j} \otimes \dyad{0}\dyad{0}{1}\dyad{1}{0}_{j+1} + \dyad{1}\dyad{0}{1}_{j} \otimes\dyad{0}{1}\dyad{1}{0}\dyad{0}_{j+1})\Big) \otimes \dyad{0}_{j+2} \nn\\
    &=-\frac{\gamma}{4} \dyad{0}{1}_{j} \otimes \dyad{0}_{j+1} \otimes \dyad{0}_{j+2},
\ea\es

such that \vspace{-6pt}
\ba
    \hat{\mathcal{L}}^\tx{(\Fuks)}&[\dyad{0...0}{0...0_{j-1}1_j0_{j+1}...0}]
    =-\gamma\dyad{0...0}{0...0_{j-1}1_j0_{j+1}...0}\neq0.
    \label{eq:appss000010}
\ea

Analogously, due to the symmetry of the Lindbladian, the action of the Lindbladian on the associate hermitian conjugate state as well as the corresponding bit-flipped states and its hermitian conjugate are all non-zero---specifically,
\ba
    \hat{\mathcal{L}}^\tx{(\Fuks)}[\dyad{0...0_{j-1}1_j0_{j+1}...0}{0...0}]
    &=\qty(\hat{\mathcal{L}}^\tx{(\Fuks)}[\dyad{0...0}{0...0_{j-1}1_j0_{j+1}...0}])^\dagger
    &&=\dyad{0...0_{j-1}1_j0_{j+1}...0}{0...0},
\nn\\[2\jot]
    \hat{\mathcal{L}}^\tx{(\Fuks)}[\dyad{1...1}{1...1_{j-1}0_j1_{j+1}...1}]
    &=\hat{X}^{(N)}\hat{\mathcal{L}}[\dyad{0...0}{0...0_{j-1}1_j0_{j+1}...0}]\hat{X}^{(N)}
    &&=\dyad{1...1}{1...1_{j-1}0_j1_{j+1}...1},
\nn\\[2\jot]
    \hat{\mathcal{L}}^\tx{(\Fuks)}[\dyad{1...1_{j-1}0_j1_{j+1}...1}{1...1}]
    &=\qty(\hat{X}^{(N)}\hat{\mathcal{L}}[\dyad{0...0}{0...0_{j-1}1_j0_{j+1}...0}]\hat{X}^{(N)})^{\!\dagger}
    &&=\dyad{1...1_{j-1}0_j1_{j+1}...1}{1...1}.
\ea

Overall, it has been shown that $\hat{\mathcal{L}}^\tx{(\Fuks)}[\hat\rho]\neq0$; see Equation~\eqref{eq:appss010} such that the state $\hat\rho$ in Equation~\eqref{eq:rhossapp010} is not a steady state of the QCA --- all steady states are of the form \eqref{eq:appssfinal}.

\section{Proof That Dephasing QCA Solves the DC Task} \label{app:Dephasing}

The proof of the conservation of the number density in the Dephasing QCA (see Section~\ref{sec:dephasing}) follows in an analogous manner to the proof for the \Fuks\ QCA (see Section~\ref{sec:fuks}) by showing the conservation of the expectation value of $\hat{S}_z=\frac{1}{2}\sum_i \hat{Z}_i$ as outlined in Appendix~\ref{app:SzFuks}.
In contrast, the Dephasing Lindbladian acts on only two instead of three neighboring sides, $j$ and $j+1$, such that Equation~\eqref{eq:adjointliosum} becomes
\ba
    \qty(\hat{\mathcal{L}}^\tx{(Dephasing)})^{\!\dagger} \qty[\hat{S}_z]
    =\frac{1}{2} \sum_{j} \qty(\qty(\hat{\mathcal{L}}_{j,j+1}^\tx{(Dephasing)})^{\!\dagger}\qty[\hat{Z}_{j}]
    +\qty(\hat{\mathcal{L}}_{j,j+1}^\tx{(Dephasing)})^{\!\dagger}\qty[\hat{Z}_{j+1}]).
    \label{eq:adjointdephasing}
\ea

Plugging in the definition of the Dephasing Lindbladian in Equations~\eqref{eq:LioDephasing} and \eqref{eq:HDephasing} leads to

\ba
\qty(\hat{\mathcal{L}}_{j,j+1}^\tx{(Dephasing)})^{\!\dagger}&\qty[\hat{Z}_{j}]
    +\qty(\hat{\mathcal{L}}_{j,j+1}^\tx{(Dephasing)})^{\!\dagger}\qty[\hat{Z}_{j+1}]
    =
    -i\,\Omega \ \qty[\hat{X}_j\hat{X}_{j+1}+\hat{Y}_j\hat{Y}_{j+1} , \hat{Z}_j+\hat{Z}_{j+1}]
    \nn\\
    &+
     \sum_{k=1}^4 \qty(
    L_{k_{j,j+1}}^\dagger (\hat{Z}_j+\hat{Z}_{j+1}) \hat{L}_{k_{j,j+1}}
    -\frac{1}{2}
    \qty((\hat{Z}_j+\hat{Z}_{j+1}) \hat{L}_{k_{j,j+1}}^\dagger \hat{L}_{k_{j,j+1}} + \hat{L}_{k_{j,j+1}}^\dagger \hat{L}_{k_{j,j+1}} (\hat{Z}_j+\hat{Z}_{j+1}))).
\ea

The Hamiltonian term  is vanishing $\forall j$ as
\vspace{-12pt}

\ba
    \qty[\hat{X}_j\hat{X}_{j+1}+\hat{Y}_j\hat{Y}_{j+1} \; , \, \hat{Z}_j+\hat{Z}_{j+1}] 
    &= \qty[\hat{X}_j,\hat{Z}_j]\hat{X}_{j+1} + \qty[\hat{Y}_j,\hat{Z}_j]\hat{Y}_{j+1} + \hat{X}_j\qty[\hat{X}_{j+1},\hat{Z}_{j+1}] + \hat{Y}_j\qty[\hat{Y}_{j+1},\hat{Z}_{j+1}]\nn\\
    &= -2i\hat{Y}_j\hat{X}_{j+1} +2i\hat{X}_j\hat{Y}_{j+1} -2i\hat{X}_j\hat{Y}_{j+1} +2i\hat{Y}_j\hat{X}_{j+1} \nn\\
    &= -2i \qty(\hat{Y}_j\hat{X}_{j+1} - \hat{X}_j\hat{Y}_{j+1} + \hat{X}_j\hat{Y}_{j+1} - \hat{Y}_j\hat{X}_{j+1}) = 0,
\ea
due to the commutation relations of the Pauli operators, $\qty[\hat\sigma^i,\hat\sigma^j]=2i\varepsilon^{ijk}\hat\sigma^k \ \forall \ \hat\sigma^l \in \big\{\hat{X},\hat{Y},\hat{Z}\big\}$ with $l\in\{i,j,k\}$ where $\varepsilon^{ijk}$ the Levi--Civita symbol.

Next, the dissipator is vanishing too because the summands for all four jump operators in Equation~\eqref{eq:jumpopsDephasing1}--\eqref{eq:jumpopsDephasing4} are zero $\forall j$. That is, the first jump operator $\hat{L}_{1_{j,j+1}} = \hat{P}_{0_j} \hat{P}_{0_{j+1}}$ yields
\ba
    &\hat{L}_{1_{j,j+1}}^\dagger \qty(\hat{Z}_j+\hat{Z}_{j+1}) \hat{L}_{1_{j,j+1}}
    -\frac{1}{2}
    \qty(\qty(\hat{Z}_j+\hat{Z}_{j+1}) \hat{L}_{1_{j,j+1}}^\dagger \hat{L}_{1_{j,j+1}} + \hat{L}_{1_{j,j+1}}^\dagger \hat{L}_{1_{j,j+1}} \qty(\hat{Z}_j+\hat{Z}_{j+1})) \nn\\
    &= \hat{P}_{0_j} \hat{P}_{0_{j+1}} \qty(\hat{Z}_j+\hat{Z}_{j+1}) \hat{P}_{0_j} \hat{P}_{0_{j+1}}
    -\frac{1}{2}
    \qty(\qty(\hat{Z}_j+\hat{Z}_{j+1}) \hat{P}_{0_j} \hat{P}_{0_{j+1}} + \hat{P}_{0_j} \hat{P}_{0_{j+1}} \qty(\hat{Z}_j+\hat{Z}_{j+1})) \nn\\
    &= \hat{P}_{0_j} \hat{P}_{0_{j+1}}
    -\frac{1}{2}
    \qty(\hat{P}_{0_j} \hat{P}_{0_{j+1}} + \hat{P}_{0_j} \hat{P}_{0_{j+1}})  = 0
\ea

\noindent
with $\hat{P}_0=\dyad{0}$ and $\hat{P}_0 \hat{Z} = \hat{Z} \hat{P}_0 = \hat{P}_0 \hat{Z} \hat{P}_0 = \hat{P}_0$;
and analogously, the term of the fourth jump operator $\hat{L}_{4_{j,j+1}} = \hat{P}_{1_j} \hat{P}_{1_{j+1}}$ leads to:
\ba
    &\hat{L}_{4_{j,j+1}}^\dagger (\hat{Z}_j+\hat{Z}_{j+1}) \hat{L}_{4_{j,j+1}}
    -\frac{1}{2}
    \qty((\hat{Z}_j+\hat{Z}_{j+1}) \hat{L}_{4_{j,j+1}}^\dagger \hat{L}_{4_{j,j+1}} + \hat{L}_{4_{j,j+1}}^\dagger \hat{L}_{4_{j,j+1}} (\hat{Z}_j+\hat{Z}_{j+1})) \nn\\
    &= -\hat{P}_{1_j} \hat{P}_{1_{j+1}}
    -\frac{1}{2}
    \qty(-\hat{P}_{1_j} \hat{P}_{1_{j+1}} - \hat{P}_{1_j} \hat{P}_{1_{j+1}}) \nn\\
    &= 0
\ea

\noindent
with $\hat{P}_1=\dyad{1}$ and $\hat{P}_1 \hat{Z} = \hat{Z} \hat{P}_1 = \hat{P}_1 \hat{Z} \hat{P}_1 = -\hat{P}_1$.
Last, the summands of the second and third jump operators $\hat{L}_{2_{j,j+1}} = \dyad{\psi^+}_{j,j+1}$ and $\hat{L}_{3_{j,j+1}} = \dyad{\psi^-}_{j,j+1}$ are vanishing as
\ba
    &\hat{L}_{2,3_{j,j+1}}^\dagger (\hat{Z}_j+\hat{Z}_{j+1}) \hat{L}_{2,3_{j,j+1}}
    -\frac{1}{2}
    \qty((\hat{Z}_j+\hat{Z}_{j+1}) \hat{L}_{2,3_{j,j+1}}^\dagger \hat{L}_{2,3_{j,j+1}} + \hat{L}_{2,3_{j,j+1}}^\dagger \hat{L}_{2,3_{j,j+1}} (\hat{Z}_j+\hat{Z}_{j+1})) \nn\\
    &= 0-\frac{1}{2}(0 + 0) = 0.
\ea

All in all, it has thus been shown that $\qty(\hat{\mathcal{L}}^\tx{(Dephasing)})^{\!\dagger} \qty[\hat{S}_z]=0$ according to Equation~\eqref{eq:adjointdephasing}  such that the expectation value of $\hat{S}_z$ is conserved:
\ba
    \frac{\tx d}{\tx dt}\expval{\hat{S}_z(t)} = 0.
\ea

Next, it is shown that the Dephasing QCA is therefore a density classifier due to its translation invariance. Let $\ket{\hat\rho_\tx{ss}}$ with $\hat{\mathbb{L}}\ket{\hat\rho_\tx{ss}}=0$ be a vectorized steady state of a system defined by a translationally invariant Lindbladian $\hat{\mathbb{L}}$, then $\hat{T}\hat\rho_\tx{ss}\hat{T}^\dagger \rightarrow \hat{T}\otimes \hat{T}^* \ket{\hat\rho_\tx{ss}}$, with translation operator $\hat{T}$, must also be a steady state of the same system as
\ba
    \hat{\mathbb{L}} \,\qty(\hat{T}\otimes\hat{T}^*) \ket{\hat\rho_\tx{ss}}
    &= \qty(\hat{T}\otimes\hat{T}^*) \underbrace{\qty(\hat{T} \otimes \hat{T}^*)^{\!\dagger} \; \hat{\mathbb{L}} \; \qty(\hat{T}\otimes\hat{T}^*)}_{=\hat{\mathbb{L}}} \ket{\hat\rho_\tx{ss}} \nn\\
    &= \qty(\hat{T}\otimes\hat{T}^*) \,\hat{\mathbb{L}} \,\ket{\hat\rho_\tx{ss}} = 0,
\ea
such that any single site translation of a steady state is also a steady state.

Thus, given that the Dephasing QCA conserves $\hat{S_z}$ and is translationally invariant, it is shown to be a density classifier and solves the DC task.

\section{\Fates\ QCA} \label{sec:fates}
Analogous to the \Fuks\ CA, the \Fates\ rule is also a radius-one probabilistic CA. It is given by the traffic-majority (TM) rule studied in \cite{fates} and consists of a linear combination of the traffic rule 184 with probability $p\in[0,1]$ and the majority rule 232 with probability $1-p$:
\ba
    \mathbb{\hat S}^\tx{(\Fates)}
    \, =
    \ p \underbrace{\ \mathbb{\hat S}^{(184)}}_\tx{traffic rule}
     \, + \ \, (1-p) \underbrace{\ \ \mathbb{\hat S}^{(232)}}_\tx{majority rule},
\label{eq:cafates}
\ea

\begin{figure}[h]
    \includegraphics[width=.55\columnwidth]{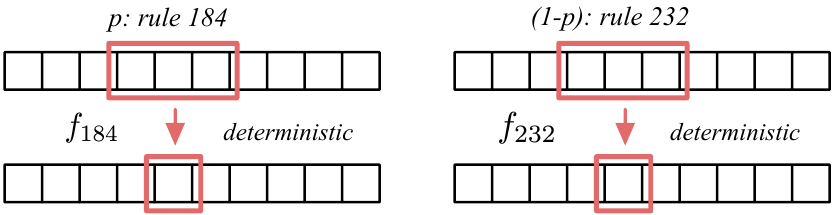}
    \caption{Illustration of the dynamics of the \Fates\ QCA, which applies a stochastic combination of the deterministic elementary CA rules 184 and 232 with probability $p$ or $1-p$, respectively.}
    \label{fig:fates}
\end{figure}

That is, the same map, either $\mathbb{\hat S}^{(184)}$ or $\mathbb{\hat S}^{(232)}$, is applied to all cells in one time step (see Figure~\ref{fig:fates}) with the corresponding transition probabilities shown in Table~\ref{tab:fates}.

\begin{table}[h]
\begingroup
\setlength{\tabcolsep}{10pt} 
\renewcommand{\arraystretch}{1.5} 
\begin{center}
	\begin{tabular}{c||c|c||c|c}
 &\multicolumn{2}{c||}{CA 184}& \multicolumn{2}{c}{CA 232}\\\hline
    neighborhood &probability & Kraus operators & probability & Kraus operators \\\hline\hline
        \multirow{ 2}{*}{00} &
        $p_{000} = 0$ & amp.~damping & $p_{000} = 0$ & amp.~damping\\
        & $p_{010} = 0$ & $\Big\{\hat P_0,\hat\sigma^-\Big\}$ & $p_{010} = 0$ & $\Big\{\hat P_0,\hat\sigma^-\Big\}$\\\hline
        \multirow{ 2}{*}{01} & $p_{001} = 0$ & identity channel & $p_{001} = 0$ & identity channel\\
        & $p_{011} = 1$ & $\big\{\hat{\mathds{1}}\big\}$ & $p_{011} = 1$ & $\big\{\hat{\mathds{1}}\big\}$\\\hline
        \multirow{ 2}{*}{10} & $p_{100} = 1$ & bit-flip & $p_{100} = 0$ &  identity channel\\
         &$p_{110} = 0$ & $\Big\{\hat{X}\Big\}$ & $p_{110} = 1$ & $\Big\{\hat{\mathds{1}}\Big\}$ \\\hline
        \multirow{ 2}{*}{11} &$p_{101} = 1$ & amp.~pumping & $p_{101} = 1$ & amp.~pumping\\
        & $p_{111} = 1$ & $\Big\{\hat P_1,\hat\sigma^+\Big\}$ & $p_{111} = 1$ & $\Big\{\hat P_1,\hat\sigma^+\Big\}$
	\end{tabular}
    \end{center}
	\caption{\Fates\ QCA. 
 Second and fourth column: transition probabilities $p_{acb}$ representing the likelihood of the state transition $\ket{acb}\rightarrow\ket{a1b}$, with $a,b,c\in\{0,1\}$ $\forall \ p\in[0,1\big]$. Third and fifth column: set of Kraus operators of the associated quantum channels acting on the center site $j$.}
\label{tab:fates}
\endgroup
\end{table}

The map's fixed point is the all-zero state if the initial state exhibits a number density less than $\frac{1}{2}$, the all-one state if the initial number density is greater than $\frac{1}{2}$, or random for an equal initial number of zero and one states. However, the desired fixed point is reached only within a certain error threshold depending on the probability $p$ and the structure of the input state, and does not solve the majority voting problem with complete accuracy; see the discussion in \cite{fates}.

Translating the CA rule in Table~\ref{tab:fates} in the Kraus operator formalism, one would, analogous to the \Fuks\ QCA, implement a complete amplitude damping (pumping) channel for the $\dyad{0}_{j-1}\otimes\dyad{0}_{j+1}$ ($\dyad{1}_{j-1}\otimes\dyad{1}_{j+1}$) neighborhood, apply the identity operation in the case of the $\dyad{0}_{j-1}\otimes\dyad{1}_{j+1}$ state, and a stochastic bit-flip in the case of the $\dyad{1}_{j-1}\otimes\dyad{0}_{j+1}$ neighborhood, as this is the only neighborhood that differentiates between the CA rules 184 and 232.
The corresponding sets of Kraus operators are the same for both rules, except in case of the 10 neighborhood:
\bs
\ba
    \hat K_{0}^{(00)} &= \dyad{0},
    &&\hat K_{1}^{(00)} = \hat\sigma^-, \\
    \hat K_{0}^{(01)} &= \hat{\mathds{1}},\\
    \hat K_{0}^{(10)} &= 
    \begin{cases}
        \hat{X} &\tx{for rule 184}\\
        \hat{\mathds{1}} &\tx{for rule 232}\\
    \end{cases},\label{eq:krausX}\\
    \hat K_{0}^{(11)} &= \dyad{1},
    &&\hat K_{1}^{(11)} = \hat\sigma^+,
\ea
\label{eq:krausfates}\es

\begin{figure}[h]
    \includegraphics[scale=0.5]{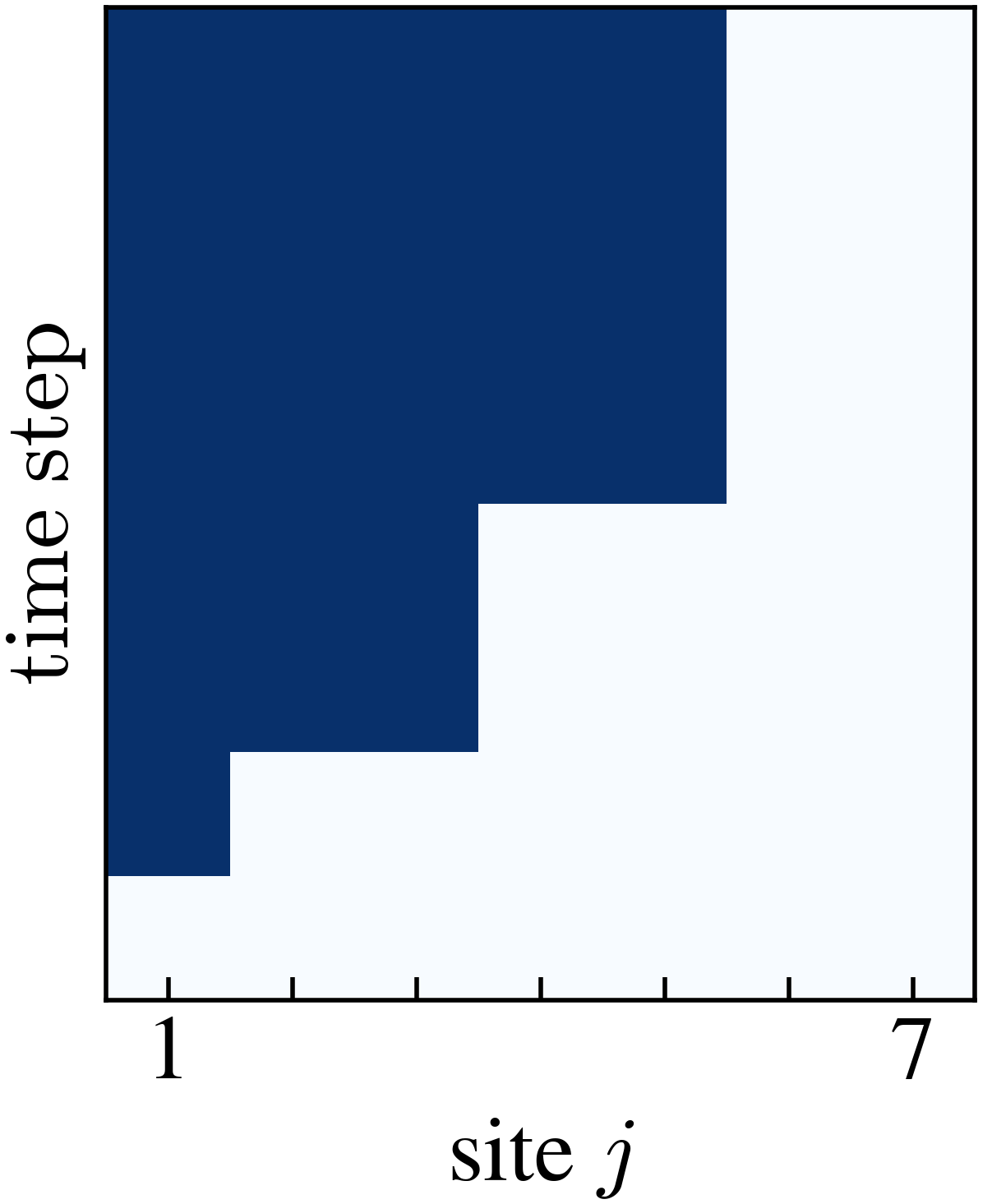}
    \caption{Application of the \Fates\ rule using an even/odd partitioning scheme with $p=\frac{1}{2}$. Starting from the state $\ket{1111100}$ with $N=7$, the system does not evolve into the desired majority steady state $\ket{1}^{\otimes N}$. One time step corresponds to one update of the QCA, i.e.,~updating all even and all odd lattice sites sequentially.}
    \label{fig:fail}

\end{figure}

These define the superoperator in Equation~\eqref{eq:s} according to Equation~\eqref{eq:kraus}.
However, this map does not lead to the desired steady state $\dyad{0...0}$ ($\dyad{1...1}$) for an initial number density less (greater) than $\frac{1}{2}$, and is therefore not further inspected.
Figure~\ref{fig:fail} shows an example in which the mapping fails, using even/odd partitioning scheme for each rule with $p=\frac{1}{2}$. Changing the value of $p$ does not improve reaching the correct steady state; rather, it affects the speeding up or slowing down of convergence.

\section{\texorpdfstring{Proof of the Properties of \boldmath{$\hat{\mathbb{A}}$} and the Dynamics of the Majority Voting Lindbladian \boldmath{$\hat{\mathbb{L}}^{\text{(ML)}}$} }{}}
\subsection{\texorpdfstring{Proof of the Properties of $\hat{\mathbb{A}}$ }{}}\label{app:MV_A}
In the following, the properties of $\hat{\mathbb{A}}$ are proved; see Section~\ref{sec:mv}.
The first property is easily demonstrated by applying operator $\mathbb{A}$ to a state featuring a cluster of $\ket{1}$s, observing that the action of $\hat{\mathbb{A}}^{m_a}$ spreads the $\ket{1}$ states out along the chain. See the below example:\vspace{-6pt}
\begin{align*}
	\ket{000011111110000} & \overset{\hat{\mathbb{A}}}{\longrightarrow} |000010111111000\rangle \\
	& \overset{\hat{\mathbb{A}}^2}{\longrightarrow} |000010101111100\rangle \\
	& \overset{\hat{\mathbb{A}}^3}{\longrightarrow} |000010101011110\rangle \\
	& \overset{\hat{\mathbb{A}}^4}{\longrightarrow} |000010101010111\rangle \\
	& \overset{\hat{\mathbb{A}}^5}{\longrightarrow} |010010101010101\rangle. \\
\end{align*}

\vspace{-23pt}
Next, $\qty[\hat{\mathbb{S}}_z,\hat{\mathbb{A}}]=0$ is derived, where $\hat{\mathbb{S}}_z$ is the vectorized form of $\hat{S}_z=\frac{1}{2}\sum_j \hat{Z}_j$. Defining $\hat{\mathbb{A}} 
=\prod_j \hat{\mathbb{A}}_j$ with $\hat{\mathbb{A}}_j = \hat{\mathbb{K}}_{0_j}+ \hat{\mathbb{K}}_{1_j}$ and $\hat{\mathbb{K}}_{\mu_j}=\hat{K}_{\mu_j}\otimes \hat{K}_{\mu_j}^{*}\ \forall \mu\in\{0,1\}$ yields  
\ba
    \qty[\hat{\mathbb{S}}_{z},\hat{\mathbb{A}}_j]
    &=\frac{1}{2}\qty[\hat{\mathbb{Z}}_{j-1}+\hat{\mathbb{Z}}_{j}+\hat{\mathbb{Z}}_{j+1}, \hat{\mathbb{K}}_{0_j} +\hat{\mathbb{K}}_{1_j}] \nn\\[2\jot] 
    &=
    \qty[\hat{Z}_{j-1},\hat{K}_{1_j}]+\qty[\hat{Z}_{j},\hat{K}_{1_j}]+[Z_{j+1},\hat{K}_{1_j}]+\qty[\hat{Z}_{j-1},\hat{K}_{0_j}]+\qty[Z_{j},\hat{K}_{0_j}]+\qty[\hat{Z}_{j+1},\hat{K}_{0_j}]
    =0,
\label{eq:commutator}
\ea
where, in the first step, we exploit the commutativity of operators on different sites, and the property $[A \otimes B, C \otimes D] = (AC) \otimes (BD) - (CA) \otimes (DB)$ is used, abandoning the vectorized form. In the second step, the first four summands are vanishing since $\qty[\hat{Z},\dyad{0}]=0=\qty[\hat{Z},\dyad{1}]$, and the last two summands cancel each other out as $\qty[\hat{Z}_{j},\hat{K}_{0_j}]=-\qty[\hat{Z}_{j+1},\hat{K}_{0_j}]=2\hat{K}_{0_j}$. Thus, taking the product over all lattice sites $j$ into account, $\hat{\mathbb{A}} 
=\prod_j \hat{\mathbb{A}}_j$:
\ba
	\qty[\hat{\mathbb{S}}_{z},\hat{\mathbb{A}}] = \underbrace{\qty[\hat{\mathbb{S}}_{z}, \hat{\mathbb{A}}_1]}_{=0} \prod_{j=2}^{N}\hat{\mathbb{A}}_j + 
	\sum_{i=1}^{N-2} \qty(\prod_{j=1}^{i} \hat{\mathbb{A}}_j) \underbrace{\qty[\hat{\mathbb{S}}_{z}, \hat{\mathbb{A}}_{i+1} ]}_{=0}  \qty( \prod_{l=i+2}^{N} \hat{\mathbb{A}}_l ) + \qty(\prod_{j=1}^{N-1}\hat{\mathbb{A}}_j ) \underbrace{\qty[\hat{\mathbb{S}}_{z}, \hat{\mathbb{A}}_N]}_{=0}  = 0 
\ea
using the property $[C, A \cdot B] = [C, A] \cdot B + A \cdot [C, B]$, where each individual commutator is vanishing due to Equation~\eqref{eq:commutator}.

\subsection{\texorpdfstring{Majority Voting Lindbladian $\hat{\mathbb{L}}^{\text{(ML)}}$ by Using a Machine Learning Approach}{}}
\label{app:LMV}
 The supervised machine learning approach is used to find an appropriate Lindbladian evolution exhibiting steady states $|0\rangle^{\otimes N}$ and $|1\rangle^{\otimes N}$ that represent the corresponding majority state of the initial state. The ansatz for the set of jump operators is to take into account all four possible neighboring state combinations $\dyad{\alpha}_{j-1}\otimes\dyad{\beta}_{j+1}$, and both the amplitude raising and lowering operators, $\hat{\sigma}^+_j$ and $\hat{\sigma}^-_j$, acting on the center site. That is, given the Lindbladian in vectorized form\vspace{-6pt}
\ba
 \hat{\mathbb{L}}^\tx{(ML)}(\vec{w})= \sum_{j=1}^{N}\sum_{k=1}^{8} \qty(  \hat L_{k_j} \otimes \hat L_{k_j}^*
 -\frac{1}{2} \qty(\hat L_{k_j}^{\dagger}\hat L_{k_j} \otimes  \hat{\mathds{1}}+  \hat{\mathds{1}} \otimes \hat L_{k_j}^{\dagger}\hat L_{k_j}) ),
\label{eq:LioMV}
\ea \vspace{-6pt}
where $\vec{w}=(w_1,...,w_8)$ and the eight considered jump operators are
\bs\ba
   \hat{L}_{1_j} &= \sqrt{w_1} \dyad{0}_{j-1} \otimes \hat{\sigma}_j^+ \otimes \dyad{0}_{j+1},\\
   \hat{L}_{2_j} &= \sqrt{w_2} \dyad{0}_{j-1} \otimes \hat{\sigma}_j^- \otimes \dyad{0}_{j+1},\\
   \hat L_{3_j} &= \sqrt{w_3} \dyad{0}_{j-1} \otimes \hat{\sigma}_j^+ \otimes \dyad{1}_{j+1}, \\
   \hat L_{4_j} &= \sqrt{w_4} \dyad{0}_{j-1} \otimes \hat{\sigma}_j^- \otimes \dyad{1}_{j+1},\\
  \hat L_{5_j} &= \sqrt{w_5} \dyad{1}_{j-1} \otimes \hat{\sigma}_j^+ \otimes \dyad{0}_{j+1}, \\
  \hat L_{6_j} &= \sqrt{w_6} \dyad{1}_{j-1} \otimes \hat{\sigma}_j^- \otimes \dyad{0}_{j+1}, \\
  \hat L_{7_j} &= \sqrt{w_7} \dyad{1}_{j-1} \otimes \hat{\sigma}_j^+ \otimes \dyad{1}_{j+1}, \\
  \hat L_{8_j} &= \sqrt{w_8} \dyad{1}_{j-1} \otimes \hat{\sigma}_j^- \otimes \dyad{1}_{j+1}.
\ea
\label{eq:jumpops_ml}\es
Since the states $|0\rangle^{\otimes N}$ and $|1\rangle^{\otimes N}$ are the desired steady states, the weights $w_1$ and $w_8$ are set to zero because their corresponding jump operators would transform the state $\ket{0_{j-1} 0_{j} 0_{j+1}}$ into the state $\ket{0_{j-1} 1_{j} 0_{j+1}}$, and $\ket{1_{j-1} 1_{j} 1_{j+1}}$ into $\ket{1_{j-1} 0_{j} 1_{j+1}}$, respectively.
The remaining decay rates $(w_2, ..., w_7)$ are determined by the ML algorithm using the following training~set:\vspace{-12pt}
\begin{align*}
	\vec{x}_1 &= [0, 0, 0, 0]  \quad\longrightarrow y_1=0, \\
	\vec{x}_2 &= [1, 0, 0, 0]  \quad\longrightarrow y_2=0, \\
	\vec{x}_3 &= [1, 0, 1, 1]  \quad\longrightarrow y_3=1, \\
	\vec{x}_4 &= [1, 0, 0, 0, 0]  \longrightarrow y_4=0, \\
	\vec{x}_5 &= [1, 1, 0, 0, 0]  \longrightarrow y_5=0, \\
	\vec{x}_6 &= [1, 0, 1, 0, 0]  \longrightarrow y_6=0, \\
	\vec{x}_7 &= [1, 1, 0, 1, 1]  \longrightarrow y_7=1, \\
	\vec{x}_8 &= [1, 1, 1, 0, 0]  \longrightarrow y_8=1, \\
	\vec{x}_9 &= [1, 0, 1, 1, 0]  \longrightarrow y_9=1, \\
	\vec{x}_{10} &= [1, 0, 1, 0, 1]  \longrightarrow y_{10}=1, \\
	\vec{x}_{11} &= [1, 1, 1, 1, 1]  \longrightarrow y_{11}=1,
\end{align*} 
for system sizes $N\in\{4,5\}$, where $X_{\text{train}}=\{\vec{x}_1,...,\vec{x}_{11} \}$, $Y_{\text{train}}=\{y_1,...,y_{11} \}$, and $y_i$ is the density classification output of the state described by $\vec{x}_i$.
The cost function is defined by
\ba
    C(\vec{w})=\sum_{i=0}^{11} (-1)^{1-y_i}f_0(\vec{w},\vec{x_i}) +(-1)^{y_i}f_1(\vec{w},\vec{x_i}),
    \label{eq:cost}
\ea
where \vspace{6pt}
\bs
\ba
    f_0(\vec{w},\vec{x_i})= \mel**{\vec{0}\,} {e^{\hat{\mathbb{L}}(\vec{w})\tau}}{\,\vec{x}_i},\\
    f_1(\vec{w},\vec{x_i})= \mel**{\vec{1}\,}{e^{\hat{\mathbb{L}}(\vec{w})\tau}}{\,\vec{x}_i},
\ea\es
with $\ket{\vec{0}}$ and $\ket{\vec{1}}$ labeling the vectorized forms of the states $\ket{0}^{\otimes N}$ and $\ket{1}^{\otimes N}$, respectively.
Note that each summand of the cost function in Equation~\eqref{eq:cost} equals $-1$ for every state that is well classified such that $C(\vec{w})=-11$ if the system solves the majority voting problem for every initial state. The goal is to minimize the cost function $C(\vec{w})$ over $\vec{w}$, setting the parameter $\tau=10\,N^2$ and letting the weights vary in the range $[0,1]$. For finding the global minimum, the basin-hopping algorithm \cite{basinhopping1,basinhopping2} with the L-BFGS-B local minimizer is used as implemented in the Python library SciPy \cite{scipy}. Once a solution is found, it is possible to achieve an improvement in convergence time by appropriately scaling the weights $\vec{w}$. However, none of the solutions $\vec{w}$ of the cost function reaches the expected value of $-11$ if all the states present in the training set are well classified. This numerical evidence seems to suggest that a global solution capable of classifying every initial state and using a Lindbladian of the form \eqref{eq:LioMV} does not exist.
Furthermore, for many of the found solutions, the cost function reaches a value close to $-9$ with only one state misclassified, namely, $\vec{x}_5$. Therefore, testing the solution on the set of states $X_{\text{test}}$ randomly generated of size $N\in\{4,5,6,7\}$, it is observed that the misclassified states belong to a specific set $R$.
Given the system size $N$, $R$ is the set of states with a majority of $\ket{0}$ states, where there exist at least two neighboring $\ket{1}$ states.

\begin{figure}[h]

\centering
    \includegraphics[width=1\columnwidth]{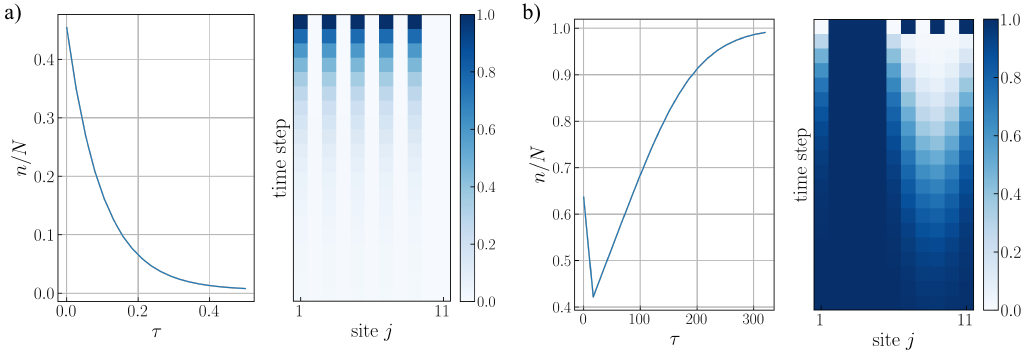}
    \caption{{Time} 
 evolution according to the Lindbladian \eqref{eq:LioMV} of the initial states (\textbf{a}) $\ket{10101010100}$ and (\textbf{b}) $\ket{01111010101}$ with $N=11$.
    On the left of each image, the normalized density $n/N$ is shown as a function of time $\tau$; on the right, one can see the time evolution of the state, reaching the state \linebreak (\textbf{a}) $\ket{0}^{\otimes11}$ and (\textbf{b}) $\ket{1}^{\otimes11}$, where each time step is equal to (\textbf{a}) 0.025$\,\tau$ and (\textbf{b}) 20$\,\tau$. Note that in (\textbf{b}), the density is decreasing for small $\tau$ and then monotonically increasing as observed in the plot on the right showing the time evolution of the state.}
 \label{fig:ML_qca}

\end{figure}

A numerical solution for the non-zero weights is $w_2=1.000$, $w_3=0.043$, $w_5=0.040$, and $w_7=0.075$, where the first three decimal digits are taken into account to ensure that the expectation value of $\hat{n}=\sum_j \hat{P}_{1_j}$ exceeds $0.99$. Analyzing how the Lindbladian evolves states not belonging to $R$ enables understanding its action and thus comprehending its effectiveness. As observable in illustrative examples below (see Figure~\ref{fig:ML_qca}), the action of $\hat{\mathbb{L}}^{\text{(ML)}}$ can be summarized by two contributions:
\begin{enumerate}
	\item[{$(i)$}] 
 It transforms every part of the string containing an alternation of zeros and ones into all zeros: $\ket{...010101...}\rightarrow \ket{...00000...}$;
	\item[{$(ii)$}] It transforms every cluster of ones  into a larger cluster by progressively adding ones both to the right and to the left across the entire chain: $\ket{...000111000...} \rightarrow \ket{...001111100...}\rightarrow \ket{...011111110...}$.
\end{enumerate}

\begin{figure}[h]
\includegraphics[scale=0.55]{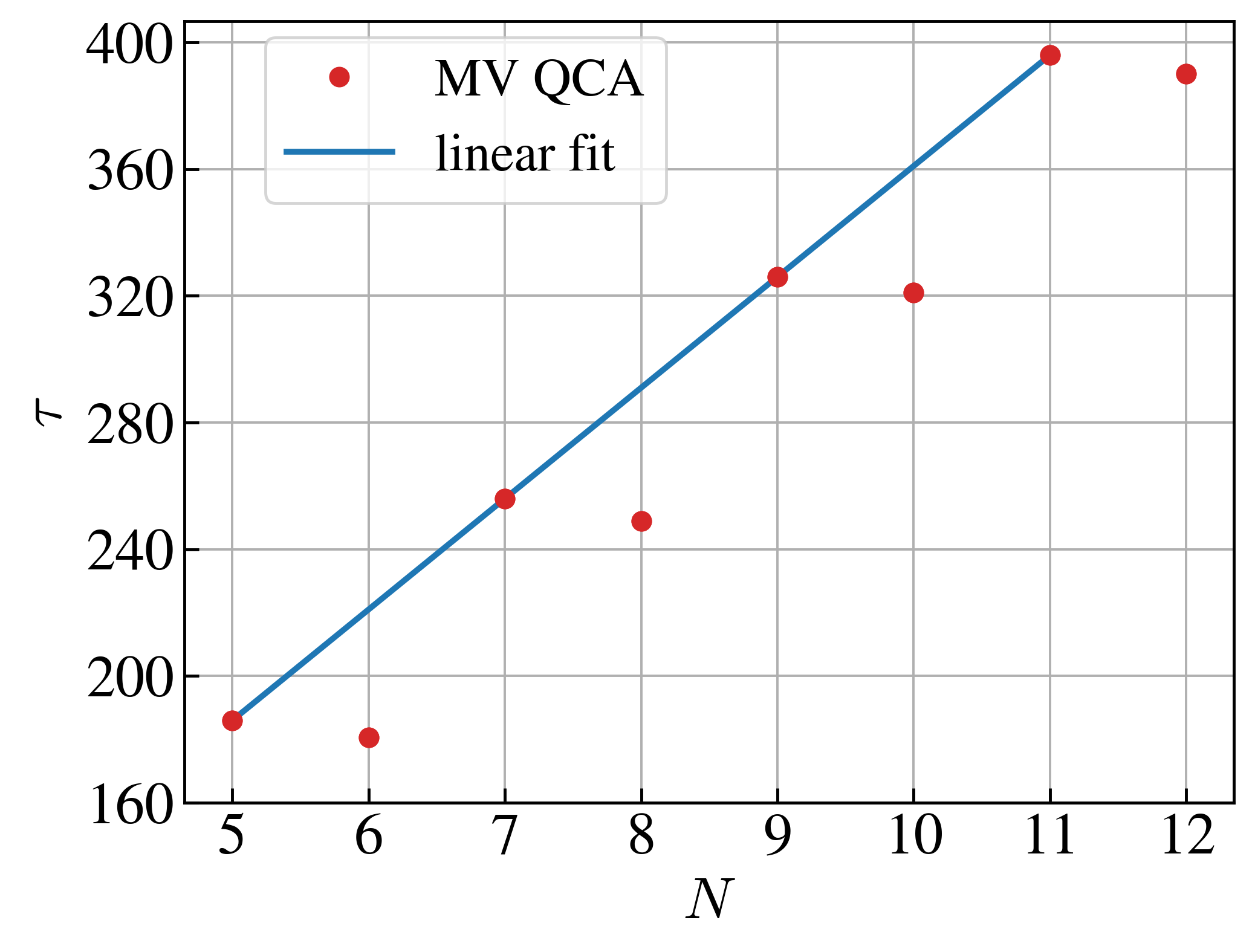}
    \caption{Time $\tau$ needed to reach the steady state of the majority voting (MV) (in the worst-case scenarios and in the sector $n>N/2$) as a function of the system size $N\in[5,12]$, computed using the QuTiP library in Python \cite{johansson2012qutip}. The linear regression fit corresponds to the data of $\tau$ associated with the odd lattice site, yielding $\tau(N)=c\times N+d$ with the parameters $c=34.992\pm 0.005$ and $d=11.03 \pm 0.04$.}
 \label{fig:ML_linear}

\end{figure}

From this, we come to realize that what we have uncovered through the ML approach is essentially equivalent to the Lindbladian in Equation~\eqref{eq:LB}, employing the conventional rule of updating the central cell.
However, evolving with this Lindbladian, the times needed to reach the final state in the worst-case scenario, are extremely higher (as it is shown in Figure~\ref{fig:ML_linear}) such that this solution has not been further explored.

A usual, we consider the steady state reached when $n/N$ exceeds $0.99$.
\section{\texorpdfstring{Proof of the Scaling of \boldmath{$\tau$} with System Size \emph{N} in Discrete-Time Evolution}{}}\label{app:proofAB}
As described in the body of the paper, the goal of transformation $\hat{\mathbb{A}}$ is to separate the clusters of $|1\rangle$s and spread them throughout the chain. When $\hat{\mathbb{A}}$ operates in the sector $n > N/2$ (where $n$ is the expectation value of $\hat{n}=\sum_j \hat{P}_{1j}$ as usual), it will never manage to completely separate the $\ket{1}$s, and a cluster will always survive. It will be thanks to the survival of this cluster that $\hat{\mathbb{B}}$ will be able to bring the resulting state to $\ket{1}^{\otimes N}$ (the action of $\hat{\mathbb{A}}$ in this sector would not be necessary). Conversely, when $\hat{\mathbb{A}}$ acts on a state with $n \leq N/2$, it will be able to completely separate the $\ket{1}$s, and $\hat{\mathbb{B}}$ will bring the state to $\ket{0}^{\otimes N}$.

In this appendix, we aim to address the following questions: using the partition scheme depicted in Figure~\ref{fig:updates} for $\hat{\mathbb{A}}$, what is the minimum number of time steps $\tau$ required to ensure that, beginning from any initial state with $n \leq N/2$, the resultant state does not contain any consecutive $\ket{1}$s? Using the partition scheme depicted in Figure~\ref{fig:updates} for $\hat{\mathbb{B}}$, what is the minimum number of time steps $\tau$ required to ensure that, after applying $\hat{\mathbb{A}}$, the resultant state is correctly classified?

To address the first question, we must first identify the worst-case scenario, which is the one requiring the greatest number of steps. Then, we need to understand how $\hat{\mathbb{A}}$, using a partition scheme, reaches the solution.

We will start by understanding how  $\hat{\mathbb{A}}$ (without any partition scheme) acts on a state having a cluster of $d$ states $|1\rangle$s: \vspace{-6pt}
\begin{align*}
&	\vert \cdots 00\overbrace{1_j1_{j+1}1_{j+2}\cdots1_{j+d}}^{d}00000 \cdots \rangle \\
&	\vert \ldots 001_j0_{j+1}\overbrace{1_{j+2}\cdots1_{j+d+1}}^{d-1}0000 \cdots \rangle \\
& \vert \ldots 001_j0_{j+1}1_{j+2}0_{j+3}\overbrace{1_{j+4}\cdots1_{j+d+2}}^{d-2}000 \cdots \rangle \\
&	\ldots \\
&\vert \ldots 001_j0_{j+1}1_{j+2}0_{j+3} 1_{j+4} \cdots 1_{j+2d-1} \cdots \rangle \\
\end{align*}

\vspace{-18pt}
So the first application of  $\hat{\mathbb{A}}$ splits the cluster by inserting a zero after the leftmost one and shifting the remaining ones by one site. It is evident, therefore, that the number of times  $\hat{\mathbb{A}}$ needs to be applied is equal to the number of zeros required to completely split the cluster, which is $d-1$.

We now demonstrate that the worst-case scenario is equivalent to splitting a single cluster with $\frac{N}{2}$ ones. Indeed, let us consider two clusters, each having $d_1$ and $d_2$ ones respectively, with $d_1$ to the left of $d_2$, separated by $l$ zeros.
If $l > d_1$, then the two clusters will be divided independently without ever interfering with each other and will be completely separated in a number of steps equal to $\max(d_1, d_2) - 1$: \vspace{-6pt}
\begin{align*}
	&	\vert \cdots 0\overbrace{1_j1_{j+1}1_{j+2}\cdots1_{j+d_1}}^{d_1}\overbrace{0000 \cdots
0}^{l} \overbrace{1_k1_{k+1}1_{k+2}\cdots1_{k+d_2}}^{d_2}00000 \cdots \rangle \\
	&	\vert \cdots 01_j0_{j+1}\overbrace{1_{j+2}\cdots1_{j+d_1+1}}^{d_1-1}\overbrace{000 \cdots
		0}^{l-1} 1_k0_{k+1}\overbrace{1_{k+2}\cdots1_{k+d_2+1}}^{d_2-1}0000 \cdots \rangle \\
	& \vert \cdots 01_j0_{j+1}1_{j+2}0_{j+3}\overbrace{\cdots1_{j+d_1+2}}^{d_1-2}\overbrace{0 \cdots 0}^{l-2}1_k0_{k+1}1_{k+2}0_{k+3}\overbrace{\cdots1_{k+d_2+2}}^{d_2-2}0 \cdots\rangle \\
	&	\ldots \\
 &	\ldots \\
	&\vert \cdots 01_j0_{j+1}1_{j+2}0_{j+3} \cdots 1_{j+2d_1-1} \overbrace{\cdots 0}^{l-d_1+1} 1_k0_{k+1}1_{k+2}0_{k+3}  \cdots 1_{k+2d_2-1} \cdots\rangle 
\end{align*}
with the index $k=j+l$. 

If $l < d_1$, then at some point, the first cluster will encounter the second one after $l$ steps, with the number of remaining ones to divide being equal to $d_1 - l$, which will be added to $d_2$. Therefore, overall, the cluster will need a number of steps equal to $l + d_1 - l + d_2 - 1 = d_1 + d_2 - 1$, which is the same number required if the initial state is composed of a single cluster with $d = d_1 + d_2$ ones, as you can see in the following~example: \vspace{6pt}
\begin{align*}
	&\begin{matrix}
	&	| \cdots 0\, 0\, \overbrace{1\, 1\, 1\, 1\, 1\, 1}^{d=6}\, 0\, 0\, 0\, 0\, 0\, 0  \cdots\rangle & | \cdots0\, 0\, \overbrace{1\, 1\, 1}^{d_1=3}\, 0\, 0\, \overbrace{1\, 1\, 1}^{d_2=3}\, 0\, 0\, 0\, 0\, 0  \cdots \rangle \\
\hat{\mathbb{A}}^{1}\longrightarrow	&	| \cdots0\, 0\, 1\, 0\, 1\, 1\, 1\, 1\, 1\, 0\, 0\, 0\, 0\, 0\, 0 \cdots\rangle & | \cdots0\, 0\, 1\, 0\, 1\, 1\, 0\, 1\, 0\, 1\, 1\, 0\, 0\, 0\, 0  \cdots\rangle \\
\hat{\mathbb{A}}^2\longrightarrow	&	| \cdots0\, 0\, 1\, 0\, 1\, 0\, 1\, 1\, 1\, 1\, 0\, 0\, 0\, 0\, 0 \cdots\rangle & | \cdots0\, 0\, 1\, 0\, 1\, 0\, 1\, 1\, 0\, 1\, 0\, 1\, 0\, 0\, 0  \cdots\rangle \\
\hat{\mathbb{A}}^3\longrightarrow	&	| \cdots0\, 0\, 1\, 0\, 1\, 0\, 1\, 0\, 1\, 1\, 1\, 0\, 0\, 0\, 0 \cdots\rangle & | \cdots0\, 0\, 1\, 0\, 1\, 0\, 1\, 0\, 1\, 1\, 0\, 1\, 0\, 0\, 0 \cdots\rangle \\
\hat{\mathbb{A}}^4\longrightarrow	&	| \cdots0\, 0\, 1\, 0\, 1\, 0\, 1\, 0\, 1\, 0\, 1\, 1\, 0\, 0\, 0 \cdots\rangle & | \cdots0\, 0\, 1\, 0\, 1\, 0\, 1\, 0\, 1\, 0\, 1\, 1\, 0\, 0\, 0  \cdots\rangle \\
\hat{\mathbb{A}}^5\longrightarrow	&	| \cdots0\, 0\, 1\, 0\, 1\, 0\, 1\, 0\, 1\, 0\, 1\, 0\, 1\, 0\, 0 \cdots\rangle & | \cdots0\, 0\, 1\, 0\, 1\, 0\, 1\, 0\, 1\, 0\, 1\, 0\, 1\, 0\, 0 \cdots\rangle \\
	\end{matrix}
\end{align*}

Generalizing to multiple clusters $d_i$ and knowing that we are in the sector $n \leq  \left\lfloor \frac{N}{2} \right\rfloor$, we have $d = \sum_i d_i \leq  \left\lfloor \frac{N}{2} \right\rfloor$, with the worst-case scenario resulting from the saturation of the~inequality.

It is important to stress that all these arguments remain valid when adopting the partition scheme for $\hat{\mathbb{A}}$: indeed, through direct examination, it is evident that what changes is the way in which the cluster is divided, but not the number of $\ket{0}$ to the right of the cluster required to complete the spreading of the $\ket{1}$s.

\begin{figure}[h]
    \includegraphics[scale=0.5]{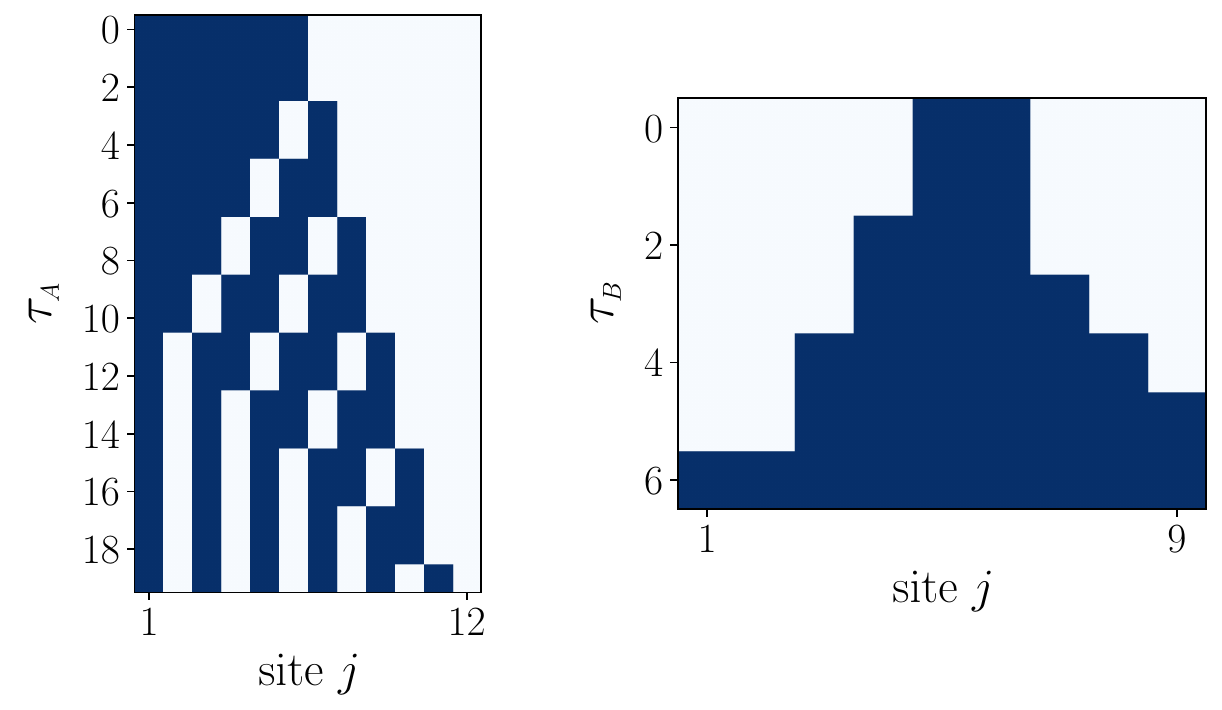}
    \caption{On the left, discrete evolution of the state $|111111000000\rangle$ by using the partitioned version of $\hat{\mathbb{A}}$. On the right, discrete evolution of the state $|000011000\rangle$ by using the partitioned version of $\hat{\mathbb{B}}$.}
 \label{fig:proofAB}

\end{figure}

Having identified the worst-case scenario, we continue our analysis by determining how many time steps $\hat{\mathbb{A}}$, using our partition scheme, needs to reach the goal.
Considering Figure~\ref{fig:proofAB} showing an example with $N=12$ in the case $N\text{mod}3=0$, it can be easily observed that, once the process of splitting begins, a zero separator is added every four layers so that the desired state is reached after $4(\left\lfloor \frac{N}{2} \right\rfloor-2) +1$. Therefore, considering that this process can, at worst, start at the third layer, we have
\ba
\tau_A = 4 \left( \left\lfloor \frac{N}{2} \right\rfloor - 2 \right) + 1 + 2 =  4  \left\lfloor \frac{N}{2} \right\rfloor -5.
\ea

\vspace{-12pt}

Now, we can address the second question regarding $\hat{\mathbb{B}}$. It is evident that the worst-case scenario occurs when the smallest possible cluster (which, for \( N\text{mod}3=0 \) and \( N \) odd, consists of two consecutive ones) needs to be expanded along the chain. In Figure~\ref{fig:proofAB}, it is illustrated how $\hat{\mathbb{B}}$, when partitioned, enlarges a cluster of three ones: once the process starts, two zeros are converted to ones every two time steps. Therefore, considering that this process can, at worst, begin at the second layer, we have
\ba
\tau_B = 2  \left( \frac{N-3}{3} \right) +1 +1 = \frac{2}{3}N. 
\ea

Finally, we obtain 
\ba
\tau=\tau_A + \tau_B =   4  \left\lfloor \frac{N}{2} \right\rfloor + \frac{2}{3}N -5.
\ea
\end{appendix}
\end{document}